\documentclass[twocolumn,showpacs,floatfix,prb]{revtex4-1}
\usepackage[english]{babel}
\usepackage{amsmath,amssymb}
\usepackage{times}
\usepackage{epsfig}
\usepackage[colorlinks,linkcolor = blue,citecolor = blue]{hyperref}
\usepackage{enumerate}
\usepackage{enumitem}
\usepackage{graphicx} 
\usepackage{hyperref}
\hypersetup{colorlinks=true}

\newcommand{\wt}{\widetilde}
\newcommand{\mc}{\mathcal}

\usepackage[usenames,dvipsnames]{xcolor}
\usepackage[normalem]{ulem}

\begin{document}
\title{Effective $S=1/2$ description of the $S=1$ chain with strong easy plane anisotropy}
\author{C. Psaroudaki$^{1,2}$, J. Herbrych$^{3,4}$, J. Karadamoglou$^{1}$, P. 
Prelov\v{s}ek$^{4,5}$, X. Zotos$^{1,2,3,6}$, and N. Papanicolaou$^{1,6}$}
\affiliation{$^1$Department of Physics, University of Crete, 71003 Heraklion, Greece}
\affiliation{$^2$Foundation for Research and Technology - Hellas, 71110 Heraklion, Greece}
\affiliation{$^3$Cretan Center for Quantum Complexity and Nanotechnology, University of Crete, 
Heraklion 71003, Greece}
\affiliation{$^4$J. Stefan Institute, SI-1000 Ljubljana, Slovenia}
\affiliation{$^5$Faculty of Mathematics and Physics, University of Ljubljana, SI-1000 Ljubljana, 
Slovenia}
\affiliation{$^6$Institute of Plasma Physics, University of Crete, 71003 Heraklion, Greece}

\date{\today}
\begin{abstract}
We present a study of the one--dimensional $S=1$ antiferromagnetic spin chain with large
easy plane anisotropy, with special emphasis on field--induced quantum phase transitions.
Temperature and magnetic field dependence of magnetization, specific heat, and thermal
conductivity is presented using a combination of numerical methods. 
In addition, the original $S=1$ model is mapped into the low--energy effective $S=1/2$ \textit{XXZ} 
Heisenberg chain, a model which is exactly solvable using the Bethe ansatz technique. 
The effectiveness of the mapping is explored, and we show that all considered quantities are
in qualitative, and in some cases quantitative, agreement. 
The thermal conductivity of the considered $S=1$ model is found to be
strongly influenced by the underlying effective description. Furthermore, we elucidate the low--lying
electron spin resonance spectrum, based on a semi--analytical Bethe ansatz calculation of the
effective $S=1/2$ model.
\end{abstract}
\pacs{75.10.Jm,75.40.-s,75.40.Gb,76.30.-v,05.60.Gg}

\hspace*{\fill}\\\hspace*{\fill} CCTP-2014-07

\maketitle

\section{Introduction}
One of the most fascinating features of a $S=1$ Heisenberg antiferromagnetic (AFM) chain is the 
occurrence of an excitation gap first suggested by Haldane [\onlinecite{haldane}]. In the presence of easy 
plane anisotropy $D$ and a magnetic field $H$ along the hard axis, the $S=1$ chain is described by 
the Hamiltonian:
\begin{equation}
\mathcal{H}=\sum_{n} \left[J\mathbf{S}_n\cdot\mathbf{S}_{n+1}+D(S_n^z)^2+H S_n^z\right]\,,
\label{s1model}
\end{equation}
where $\mathbf{S}_{n}=(S_{n}^{x},S_{n}^{y},S_{n}^{z})$.
The physical properties of the system strongly depend on the strength of anisotropy $D$. For $D=0$, 
the ground state is a singlet and the lowest excitation is a degenerate massive triplet with $S=1$. 
For positive $D$ the triplet splits into an $S^z=0$ state and a degenerate $S^z=\pm 1$ doublet with 
lower energy. When $D$ is increased, the Haldane gap is diminished until it vanishes [\onlinecite{langari}]
at some critical $D_c=0.968 J$. At this point a transition occurs, so when $D$ is further increased 
we observe the rise of a gap of different nature [\onlinecite{papanicolaou}]. 

We focus on the large--$D$ limit, where the anisotropy $D$ is much larger than the exchange 
coupling $J$. For zero magnetic field this phase is characterized by a nondegenerate ground state 
that is the direct product of states with $S^z=0$, because, due to the large anisotropy, all spins are 
forced to lie in the \textit{XY} plane. The lowest excited states can be constructed by reducing or increasing 
the azimuthal spin by one unit at a site, so that the total spin in the $z$ direction is $S^z=\pm1 
$, with a gap $\Delta_0 \sim D$. The energy momentum dispersion of these degenerate states has been 
calculated through a systematic 1/$D$ expansion carried to third order [\onlinecite{papanicolaou}]. Several 
more terms beyond the third order have become available [\onlinecite{hamer}]. 

The application of magnetic field along the $z$ direction induces a zero--temperature quantum phase 
transition at a critical field $H_1$, above which magnetization develops in the ground state and 
the spectrum of magnetic excitations becomes gapless. At this point level crossing occurs and the 
azimuthal spin of the ground state is no longer zero but increases with increasing field. 
The value of $H_1$ is defined by the gap $\Delta_0$, $H_1=\Delta_0$, for which a third--order 
approximation is given by [\onlinecite{papanicolaou1}]
\begin{equation}
H_1=D-2J+\frac{J^2}{D}+\frac{J^3}{2 D^2}\,.
\label{H1}
\end{equation}

A second transition occurs at a critical field $H_2$, above which the ground state is fully 
polarized and the gapped excitation spectrum of a magnon can be calculated exactly. The value of 
$H_2$ is defined by the lowest gap of the magnon dispersion: 
\begin{equation}
H_2=D+4 J\,.
\label{H2}
\end{equation}

A physical realization of an $S=1$ chain in the large--$D$ limit is the organic compound 
$\mbox{NiCl}_2\mbox{-SC}(\mbox{NH}_2)_2$, abbreviated as DTN, a system of weakly interacting 
chains. The field--induced quantum phase transitions (QPT) described above, as well as the 
thermodynamic and transport properties of DTN, have attracted considerable experimental and 
theoretical attention [\onlinecite{zapf},\onlinecite{zapf2014}]. Actually, DTN is considered to be the
quasi--one--dimensional limit of a 
three--dimensional (3D) system, where the exchange couplings perpendicular to the chain $J_\perp$ 
are finite but much smaller than $J$, $J_{\perp}/J\simeq 0.18$. The intermediate phase in DTN has 
been experimentally identified as a 3D \textit{XY} AFM ordered phase that can be regarded as a 
Bose-–Einstein condensate (BEC) of magnons below some critical temperature $T_N$ [\onlinecite{yin}]. The 3D 
ordering is a result of the presence of $J_\perp$, which becomes significant whenever 
the energy gap is smaller than $J_\perp$. The $S=1$ system can be mapped into a gas of semi hard
core bosons, where the 
$S^z=-1\,, 0$, and $1$ states are mapped into a state with zero, one, and two bosons per site. 
Nevertheless, it is well known that for the one--dimensional (1D) AFM, quantum fluctuations are 
strongest and only quasi--long--range phase coherence occurs, which is turned into true long range by 
the presence of weak 3D couplings.

In this paper we will concentrate on the 1D model \eqref{s1model} where quantum effects become much 
more important. We can gain a better insight into the problem if we consider the following mapping: 
when $H \to H_1$, the state with total $S^z=-1$ approaches the ground state due to the Zeeman 
energy. The idea is to project the original Hamiltonian into this low--energy subspace using a new 
$S=1/2$ representation. A mapping based on similar considerations is possible for $H \to H_2$, 
using the single magnon state and the ferromagnetic (FM) ground state. A similar analysis has been 
carried out for $S=1/2$ ladders in a magnetic field [\onlinecite{giamarchi}], but for reasons of 
completeness we give more details about the mapping in Appendix~\ref{sec:ap1}.

The original $S=1$ Hamiltonian reduces to that of the $S=1/2$ \textit{XXZ} Heisenberg AFM chain in the presence
of the magnetic field:
\begin{equation}
\wt{\mathcal{H}}=\sum_{n}\left[2J\left(\wt{S}_n^{x}\wt{S}_{n+1}^{x}+\wt{S}_n^{y}\wt{S}_{n+1}^{y}
+\Delta\wt{S}_n^{z}\wt{S}_{n+1}^{z}\right)+\wt{H}\wt{S}_n^{z}\right]\,,
\label{s12model}
\end{equation}
where $\Delta=1/2$ and $\wt{H}=-J-D+H$. Ferromagnetic order in the ground state is established when 
the magnetic field exceeds the critical value $\wt{H}_c=2J(\Delta+1)$. The whole phase can be 
described by the effective Hamiltonian \eqref{s12model}, where
\begin{enumerate}[noitemsep,nolistsep]
\item the gapped phase of model \eqref{s1model} for $H<H_1$ corresponds to the negatively FM 
ordered state of model \eqref{s12model} for $\wt{H}<-\wt{H}_c$,
\item the gapless phase of \eqref{s1model} for $H_1<H<H_2$ corresponds to gapless phase of model 
\eqref{s12model} for $-\wt{H}_c<\wt{H}<\wt{H}_c$,
\item and the FM state of model \eqref{s1model} for $H>H_2$ corresponds to the positively FM ordered 
state of model \eqref{s12model} for $\wt{H}>\wt{H}_c$.
\end{enumerate}

The obvious advantage of this mapping is that the $S=1/2$ \textit{XXZ} chain is exactly solvable. The Bethe 
ansatz technique gives explicit analytic expressions for its eigenfunction and eigenvalues, 
and the thermodynamics can be calculated through a set of nonlinear integral 
equations. Also, the complete integrability of the $S=1/2$ \textit{XXZ} quantum spin chain has some 
interesting implications on the thermal transport properties of the original $S=1$ chain.

Here we explore the effectiveness of this mapping. A first direct test can be given if we compare 
the critical fields obtained by the two models. For the first critical field, model 
\eqref{s12model} predicts $H_1=D-2 J$, which coincides with Eq.~\eqref{H1} only at first order in 
terms of $J/D$, whereas both models predict the same value for the second critical field given by 
Eq.~\eqref{H2}. This is an indication that the mapping should be more accurate close to $H_2$ 
rather than $H_1$. Throughout this paper we adopt a certain choice of parameter $D/J=4$ in our 
numerical calculations in order to be consistent with earlier work on electron spin resonance 
(ESR) theoretical analysis [\onlinecite{psaroudaki}] of model \eqref{s1model} and to obtain semiquantitative 
agreement with experimental data on DTN [\onlinecite{Zvyagin1},\onlinecite{Zvyagin2}]. Under this choice, the critical fields are $H_1/J=2.28$ 
and $H_2/J=8$ for model \eqref{s1model}, and 2 and 8 for model \eqref{s12model}, respectively.

The paper is organized as follows: In Sec.~\ref{sec:thermodynamics} we present a detailed 
calculation of the magnetization and the specific heat for both the $S=1$ model \eqref{s1model} and 
the effective $S=1/2$ model \eqref{s12model}, using a variety of numerical techniques. In 
Sec.~\ref{sec:transport} we address the calculation of dynamic correlation functions pertinent to 
the study of thermal transport in both models. Finally, in Sec.~\ref{sec:esr} we take advantage of 
the effective $S=1/2$ model in order to elucidate the field dependence of ESR in the intermediate 
phase $H_1<H<H_2$ and thus complete recent theoretical analyses [\onlinecite{psaroudaki}] carried out 
within the $S=1$ model. Our main conclusions are summarized in Sec.~\ref{sec:conclusions}, while 
some theoretical issues are relegated to two brief Appendices. 

\section{Thermodynamics}
\label{sec:thermodynamics}
This section is devoted to the calculation of the thermodynamic quantities, such as 
magnetization and the specific heat. It is important that this calculation be done for the 
original Hamiltonian directly in some numerical ways in order to test the validity of the 
approximations used while performing the mapping.

For this reason an algorithm based on the application of the renormalization group to transfer 
matrices (TMRG) is employed, where the $S=1$ quantum chain is mapped onto a two--dimensional 
classical system by a Trotter--Suzuki decomposition of the partition function [\onlinecite{Xiang}]. The main 
advantage of this method is that the thermodynamic limit can be performed exactly and results can 
be obtained with satisfactory accuracy. Moreover, a second numerical calculation is carried out on 
the basis of the finite--temperature Lanczos method (FTLM) [\onlinecite{prelovsek2013}]. Although the TMRG
results of thermodynamic quantities are considered to be more accurate, the FTLM applies also to 
the calculation of dynamic correlations such as those presented in Sec.~\ref{sec:transport} for the 
discussion of thermal transport.

According to thermodynamic Bethe ansatz (TBA), a system of nonlinear integral equations provides 
all the required information for the calculation of the free energy of model \eqref{s12model} in 
the thermodynamic limit [\onlinecite{takahashi}]. The particular value of the anisotropy 
parameter $\Delta=1/2$ is especially convenient because the calculation of thermodynamic quantities
requires a solution of only two nonlinear integral equations. More details are discussed in 
Appendix~\ref{sec:ap2}. 

\subsection{Magnetization}
In this subsection, we calculate the magnetization curve as a function of temperature and applied 
magnetic field. In a gapped spin system in the presence of external magnetic field, the Zeeman term 
is responsible for the closure of the gap and spontaneous magnetization is developed in the ground 
state. The behavior of the magnetization curve near a critical field $H_{\text{cr}}$ is nontrivial 
and depends on the model and its dimensionality. In most cases where second--order transitions 
occur, the magnetization $M$ near $H_{\text{cr}}$ behaves like
\begin{equation}
M\sim(H-H_{\text{cr}})^{1/\delta}\,.
\end{equation}

Models with the same critical exponent $\delta$ are said to belong to the same universality class independently 
of the microscopic details of the system. In general, the universality class of the model is hard 
to derive prior to a direct calculation of magnetization. For the $S=1$ Haldane chain, the critical 
exponent was found equal to $\delta=2$, a result based on an equivalent continuum limit of quantum 
chains and a mapping of the effective low--energy Lagrangian to a Bose fluid with 
$\delta$ repulsion [\onlinecite{Affleck}]. Nevertheless, a similar low--energy quantum field theory is not 
available for the large--$D$ $S=1$ chain and hence an independent calculation of the magnetization 
curve is needed. Among the models that have the same critical exponent $\delta=2$ are the $S=1/2$ 
ladders[\onlinecite{Chitra}] and the $S=1/2$ bond--alternating chain [\onlinecite{Sukai}].

The zero temperature magnetization of the $S=1/2$ \textit{XXZ} model is based on a Bethe ansatz solution of the 
Hamiltonian. More specifically, C.~N.~Yang and C.~P.~Yang [\onlinecite{Yang}] studied the ground state 
energy as a function of $\Delta$ and magnetization, and among the various results, they proved that 
$\wt{M}$ close to $\wt{H}_c$ behaves as follows
\begin{eqnarray}
\wt{M}=\frac{1}{2}-\frac{1}{\pi}\sqrt{\wt{H}_c-\wt{H}} \quad \text{for}\quad\wt{H}<\wt{H}_c\,,
\nonumber \\
\wt{M}=-\frac{1}{2}+\frac{1}{\pi}\sqrt{\wt{H}-\wt{H}_c} \quad \text{for}\quad \wt{H}>-\wt{H}_c\,.
\end{eqnarray}
Note that the dependence of $\wt{M}$ on the anisotropy constant $\Delta$ enters only through the 
critical field $\wt{H}_c=2 J(1+\Delta)$ and thus does not affect the value of the critical exponent 
$\delta=2$. However, finite temperature will cause a smoothing in the shape of the $\wt{M}(\wt{H})$ 
curve close to $\wt{H}_c$.

\begin{figure}[!ht]
\includegraphics[width=\columnwidth]{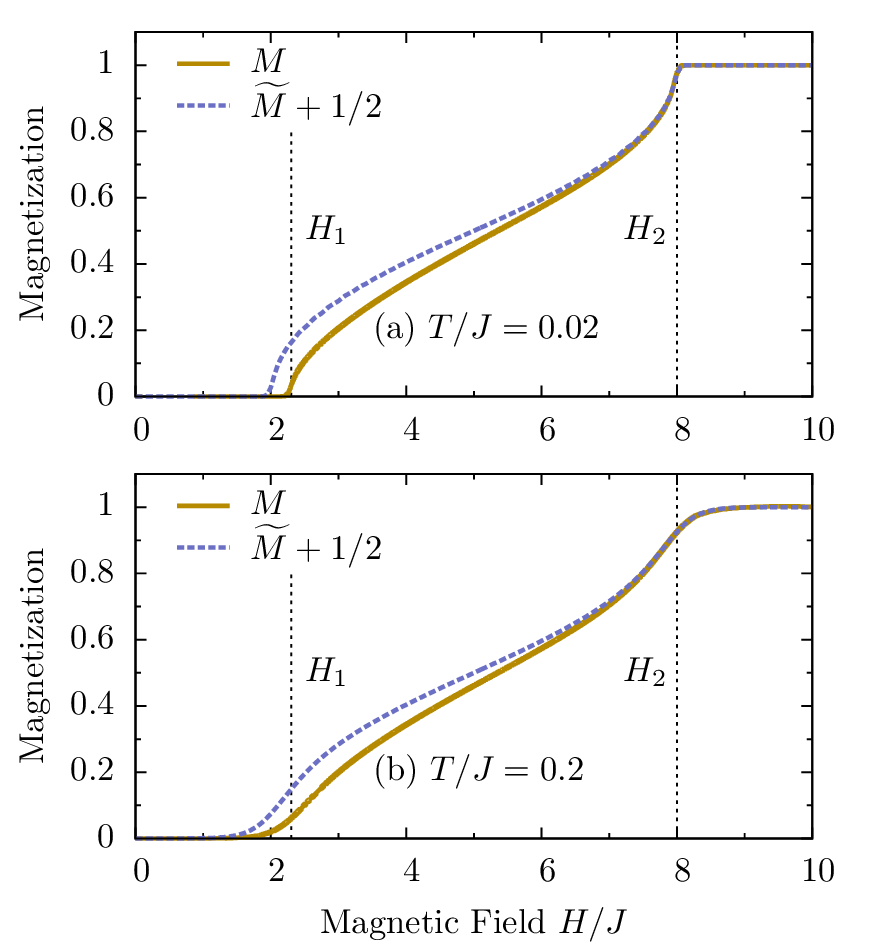}
\caption{(Color online) The magnetic field dependence of magnetization $M$ at fixed temperature 
(a) $T/J=0.02$ and (b) $T/J=0.2$. The solid line corresponds to TMRG results obtained for the $S=1$ 
large--$D$ chain and the dashed line corresponds to TBA results obtained for the $S=1/2$ \textit{XXZ} chain. 
Vertical lines indicate the location of critical fields $H_1/J=2.28$ and $H_2/J=8$. Satisfactory 
agreement between the two models is achieved, particularly close to $H_2$ where the two curves are 
indistinguishable.}
\label{magh}
\end{figure}

In Fig.~\ref{magh} we depict the magnetic field dependence of magnetization $M$ for a $S=1$ 
large--$D$ chain, superimposed with the magnetization $\wt{M}+1/2$ for the $S=1/2$ \textit{XXZ} chain for 
(a) $T/J=0.02$ and (b) $T/J=0.2$. Among the facts that become apparent are the following: (i) Temperature 
$T/J=0.02$ is considered to be low enough that the anticipated square--root behavior is evident for 
both models. The critical exponent is extracted and is found to be $\delta\simeq2$ close to $H_1$, as well as close to 
$H_2$. This foreseen result renders 
model \eqref{s1model} in the same universality class as the Haldane or $S=1/2$ \textit{XXZ} chain. (ii) As 
mentioned already, we expect that the mapping close to $H_2$ is more accurate than close to $H_1$. 
This expectation is verified by the magnetization curves close to $H_2$ which are 
indistinguishable. 

\begin{figure}[!ht]
\includegraphics[width=\columnwidth]{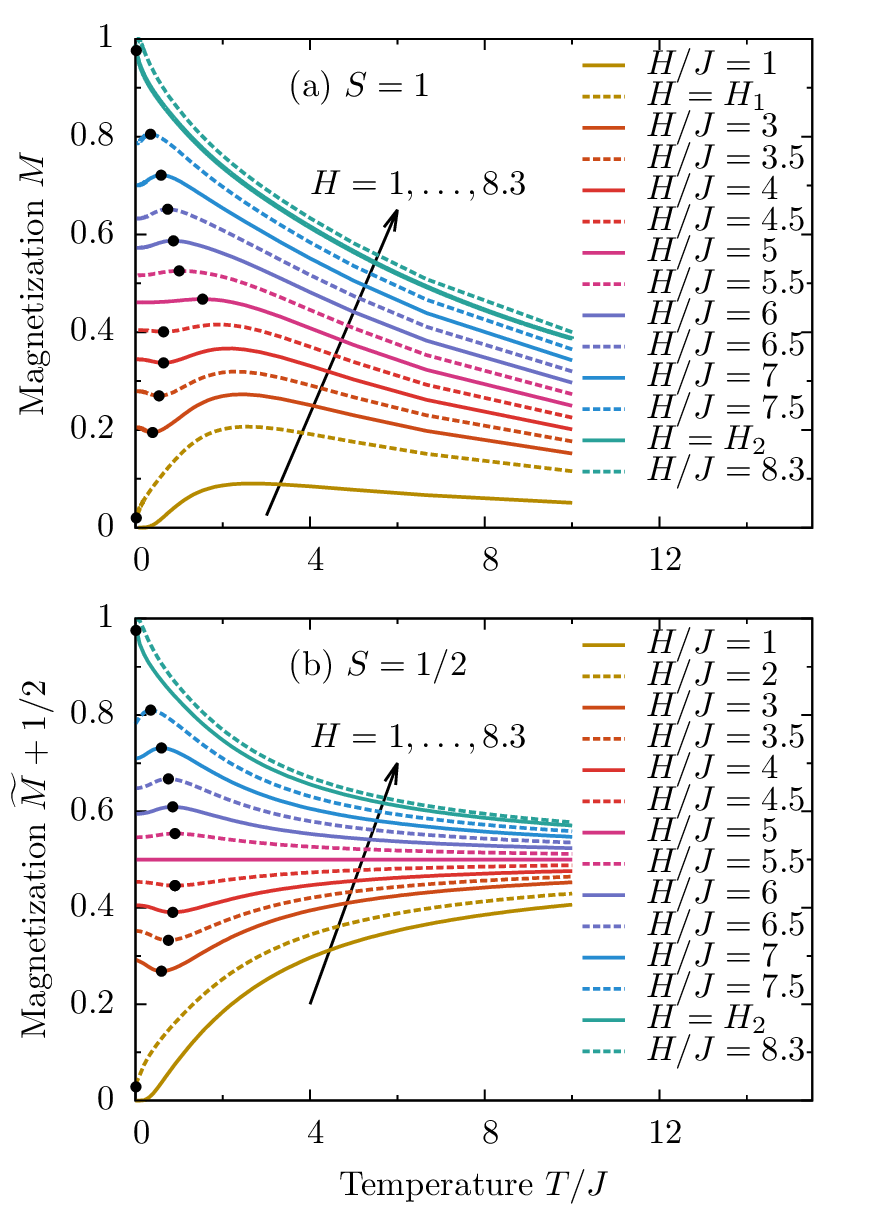}
\caption{(Color online) The temperature dependence of magnetization for (a) the $S=1$ large--$D$ model 
and (b) the $S=1/2$ \textit{XXZ} model, for various fields. Dots indicate the position of extrema that 
correspond to the Luttinger liquid crossover. $T_c$ decreases toward $T=0$ as $H$ approaches $H_1$ 
or $H_2$.}
\label{mag}
\end{figure}

Let us now focus on the temperature dependence of magnetization for a wide range of fixed magnetic 
fields, as illustrated in Fig.~\ref{mag}. For $H<H_1$, magnetization vanishes exponentially toward 
$T=0$; for $H>H_1$, a minimum appears at low temperatures that persists up to $H_m=(H_1+H_2)/2$, 
whereas maxima occur at larger magnetic fields for $H_m<H<H_2$. A further increase of the magnetic 
field will reopen the gap, and for $H>H_2$ the $M(T)$ curve decreases with increasing temperature 
and vanishes exponentially. In Fig.~\ref{mag}(a) we present the above--described behavior of $M$ and 
the position of the extrema $T_c$ is indicated by dots.

The presence of minima and maxima at low temperatures is not a surprising result, since similar 
features were found for systems of $S=1/2$ ladders [\onlinecite{Wang},\onlinecite{Wessel},\onlinecite{ruegg},\onlinecite{Bouillot}] and Haldane 
chains [\onlinecite{Maeda}], where this nontrivial behavior was interpreted as a Luttinger liquid (LL) 
crossover, with $T_c$ corresponding to the temperature below which the description of the system in 
terms of a LL is valid.

\begin{figure}[!ht]
\includegraphics[width=\columnwidth]{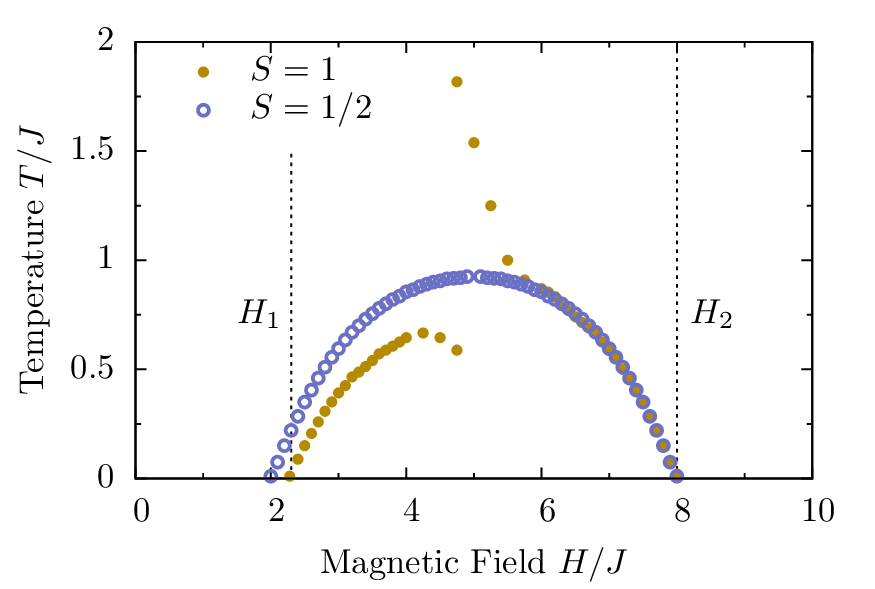}
\caption{(Color online) Magnetic phase diagram of the $S=1$ chain with a strong easy--plane anisotropy 
(full points) and of the $S=1/2$ \textit{XXZ} chain (open points). Symbols indicate the crossover into a 
finite--temperature LL regime present for both models.}
\label{phase}
\end{figure}

Here we examine this behavior in terms of the $S=1/2$ model, and in Fig.~\ref{mag}(b) we have 
plotted the temperature dependence of magnetization for the same values of magnetic field. For 
small values of temperature, magnetization behaves in a similar way, with a minimum or maximum 
being present for every value of magnetic field. Any deviations for higher temperature can be 
attributed to the missing component of the doublet. At the value $H/J=5$ ($\wt{H}=0$) the extrema 
are expected to disappear and $\wt{M}=0$ for every temperature. The position of the 
extrema is symmetric around $H/J=5$, reflecting the symmetry around $\wt{H}=0$, where every minimum 
for $\wt{H}<0$ corresponds to a maximum under the substitution $\wt{H}\rightarrow -\wt{H}$. As 
expected, this symmetry holds for the $S=1$ model only in the $D/J \gg 1$ limit. This lack of 
symmetry is easily seen in Fig.~\ref{phase}, where we present the magnetic phase diagram for both 
models with symbols marking the crossover into a low--temperature Luttinger liquid regime. 
Note that the discontinuity close to $H_m$ is an artifact of the way in which we identify the LL transition [\onlinecite{Bouillot}].

The results presented in this section, namely, the low--temperature critical exponent $\delta=2$ and 
the extrema of the $M(T)$ curve should be accessible to experimental verification. Magnetization 
measurements on DTN [\onlinecite{Paduan},\onlinecite{Weickert}] revealed a linear dependence of $M(H)$ at low 
temperatures and $M(T)$ traces at fields close to $H_1$ display a cusp--like dip that was 
attributed to the onset of 3D \textit{XY} AFM order rather than a LL crossover. Exchange couplings 
perpendicular to the chain $J_{\perp}$ play an important role in determining the dimensionality of 
DTN close to the QPT at $H_1$ and $H_2$, where the gap closes and the system behaves as 
three--dimensional. The power-law behavior of the observed phase boundary [\onlinecite{yin}] 
$H_1(T)-H_1(0)\propto T^{\alpha}$ has been identified as $\alpha=1.47\pm0.10$ consistent with the 
3D BEC universality class. We should emphasize that the phase diagram of Fig.~\ref{phase} does not 
correspond to a real phase transition, but to a crossover between different regimes with an $\alpha 
\simeq 1$ exponent, and should lie above the phase diagram of BEC or \textit{XY} AFM type.

\subsection{Specific Heat}
The magnetic field and temperature dependence of specific heat $\mc{C}_v$ is now investigated. A 
well established result [\onlinecite{papan}] is that the specific heat of the $S=1/2$ \textit{XXZ} model develops a 
characteristic double peak as a function of an applied longitudinal magnetic field at relatively 
low--$T$. This characteristic behavior cannot be explained by noninteracting magnons, where a 
single peak should be expected with its maximum at the position of the critical field.

\begin{figure}[!ht]
\includegraphics[width=\columnwidth]{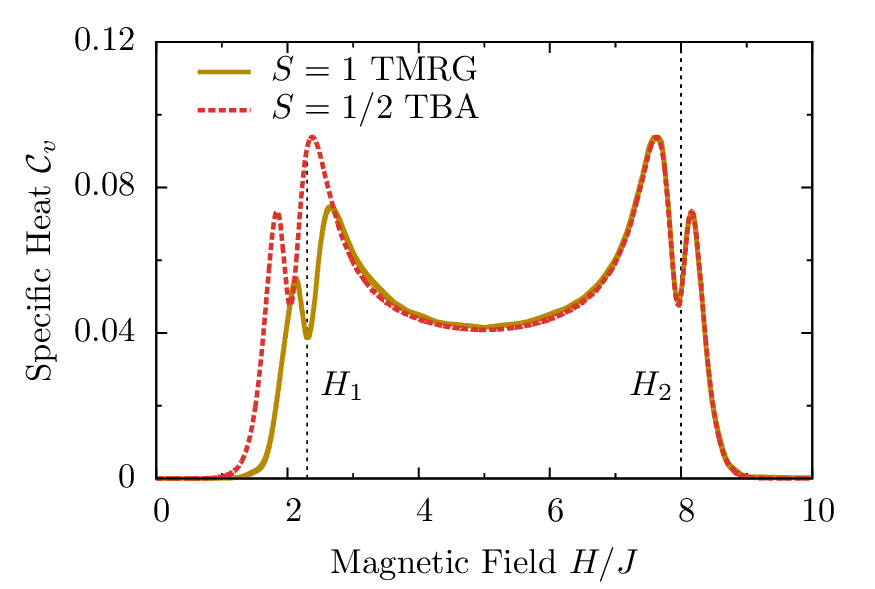}
\caption{(Color online) The magnetic field dependence of specific heat $\mc{C}_v$ at fixed 
temperature $T/J=0.1$. The solid line corresponds to TMRG results for the $S=1$ large--$D$ model and the
dashed line corresponds to TBA results of the $S=1/2$ \textit{XXZ} model.}
\label{cv_h_1}
\end{figure}

The numerical calculation of $\mc{C}_v$ for the $S=1$ large--$D$ chain reveals that the double peak 
is indeed present for adequately low temperatures. This is presented in Fig.~\ref{cv_h_1}, where 
$\mc{C}_v$ is plotted as a function of magnetic field at fixed temperature $T/J=0.1$. The position 
of the double peak is around critical fields $H_1$ and $H_2$. Note that the curve is symmetric 
around $H_m$ for the $S=1/2$ \textit{XXZ} chain due to the the spin--inversion symmetry, whereas some 
asymmetry arises for the $S=1$ large--$D$ chain which is apparent near the lower critical field 
$H_1$. 
 
\begin{figure}[!ht]
\includegraphics[width=\columnwidth]{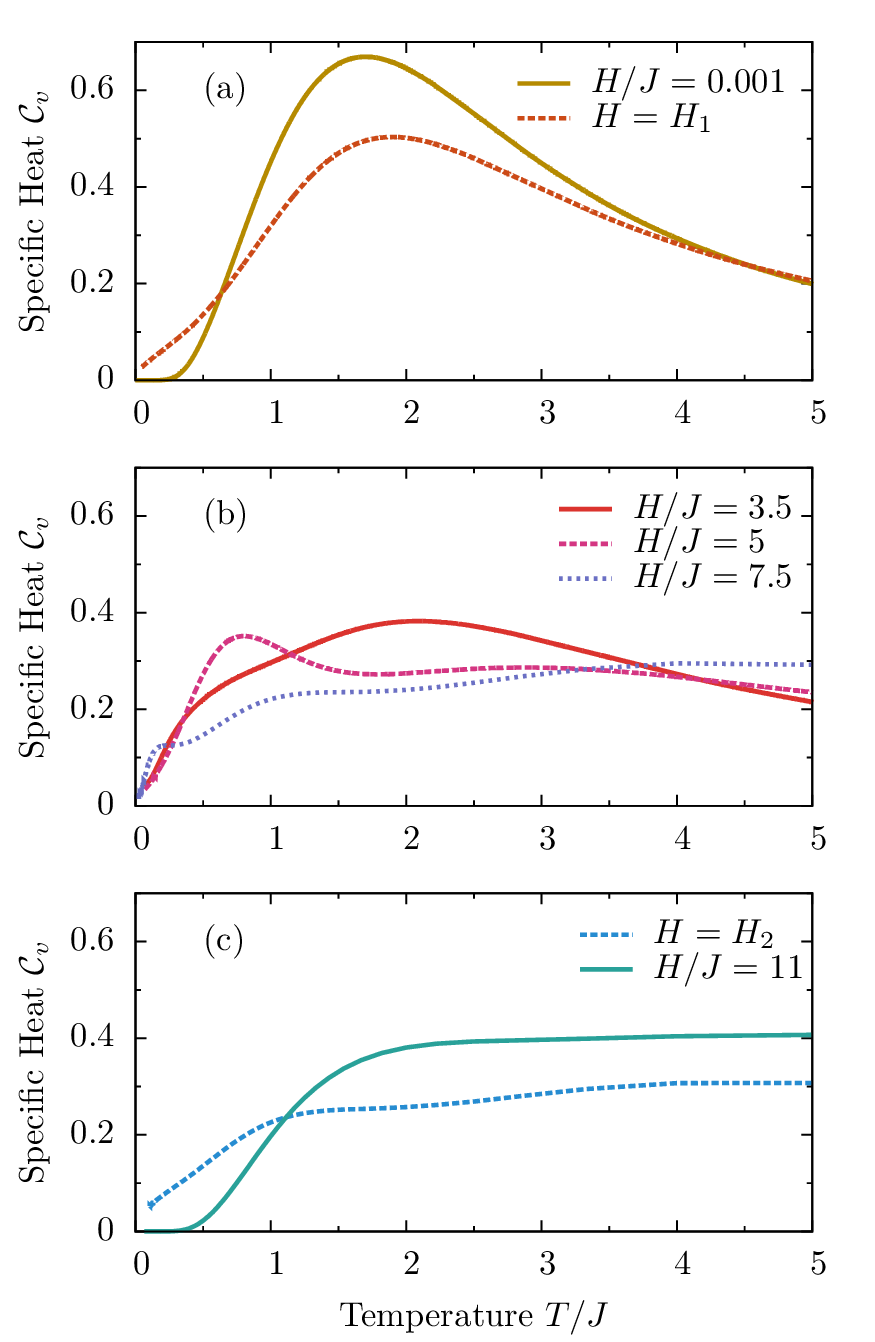}
\caption{(Color online) The temperature dependence of specific heat for various fields, calculated 
for the $S=1$ model using TMRG.}
\label{cv_t2} 
\end{figure}

The temperature dependence of specific heat is also studied at various magnetic fields, and the main 
features are depicted in Fig.~\ref{cv_t2}, calculated for the original $S=1$ model using the TMRG 
algorithm. More specifically, for $H<H_1$ specific heat decays 
exponentially at low temperatures due to the presence of the gap. The curve has a single peak which 
can be attributed to the thermal population of the $S^z=\pm1$ doublet excitations. An increase of 
$H$ will cause a decrease of the $\mathcal{C}_v$ curve. As $H \to H_1$ the gap is reduced and the 
line shape is changed, as we find linear dependence on $H$ at low--$T$. For $H_1<H<H_2$ an additional 
peak is gradually developed, below which the temperature dependence remains linear. This behavior is 
is consistent with the LL phase where specific heat scales like $\mc{C}_v/T\propto T^{d-1}$ for excitations 
with relativistic dispersion, where $d$ is the dimension. Finally, for $H>H_2$ the second peak vanishes and the 
reopening of the gap will again cause $\mathcal{C}_v$ to decay exponentially at low $T$. 

The characteristic behavior of specific heat described in this section can be fould in other models as well, 
for example, $S=1/2$ ladders. Measurements on systems of weakly coupled ladders[\onlinecite{Bouillot}] revealed 
qualitatively the same $\mc{C}_v(T)$ behavior, where the first peak in $T$ was explained as a sign 
of deviations from the LL linear regime. Moreover, the characteristic double peak of $\mc{C}_v$ as a function of magnetic 
field presented in Fig.~\ref{cv_h_1} has been found experimentally [\onlinecite{ruegg},\onlinecite{sologubenko}]. 
Note that the  $S=1/2$ ladder compounds are considered to be 
good candidates to explore effects that occur in 1D quantum systems, with the interladder coupling being 2 
orders of magnitude smaller than the intraladder couplings. 

On the contrary, the specific heat data of DTN exhibit sharp peaks as a function of $T$ and $H$, suggesting that DTN 
can partially be described as a quasi--1D system, making the inclusion of interchain couplings 
necessary in order to explain the experimental data. The low--$T$ dependence of specific heat data 
is $T^{3/2}$ at $H_1$, in agreement with the expected 3D BEC [\onlinecite{Weickert}]. In addition, the 
$\mc{C}_v(H)$ data exhibit sharp asymmetric peaks at the critical fields $H_1$ and $H_2$, an 
asymmetry that was explained in terms of mass renormalization of the elementary excitations due to 
quantum fluctuations that exist for $H\leq H_1$ and are absent for $H\geq H_2$ [\onlinecite{Kohama}]. The 
free magnon picture at any dimensionality is not sufficient to reproduce the double--peak shape. 
On the contrary, a single, rather sharp peak is predicted 
with a maximum at the critical fields. In Fig.~\ref{cv_h_1} we notice that the asymmetry in 
$\mc{C}_v$ is present for the 1Dl case as well, with the value of $\mc{C}_v$ at the 
double peak around $H_2$ being larger than the one around $H_1$. In terms of the effective mapping 
that we are discussing here, perfect symmetry is only expected in the $D/J\gg 1$ limit. 

\begin{figure}[!ht]
\includegraphics[width=\columnwidth]{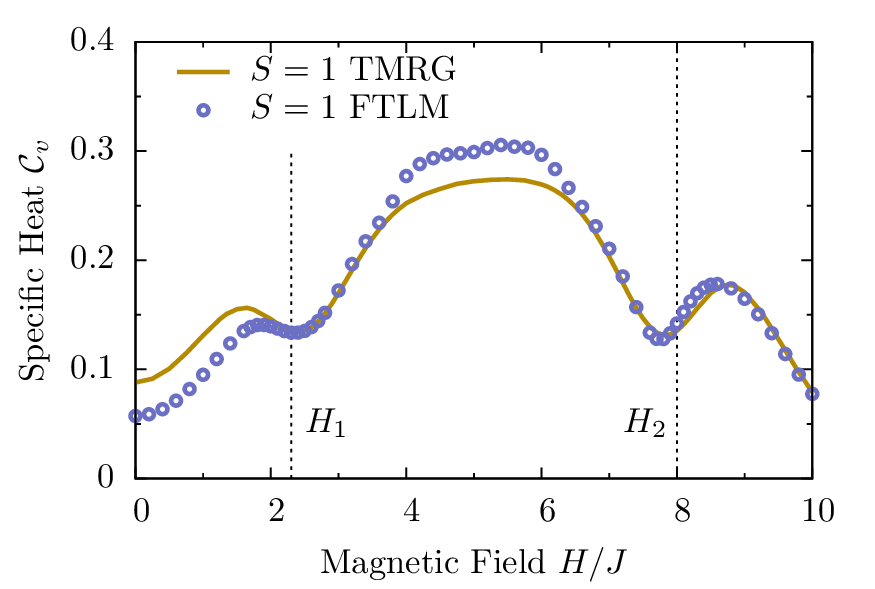}
\caption{(Color online) The magnetic field dependence of specific heat $\mc{C}_v$ at fixed 
temperature $T/J=0.5$ as calculated with TMRG (solid line) and FTLM (points) for the $S=1$ model. 
Deviations are due to finite--size effects of FTLM data.}
\label{cv_h_2}
\end{figure}

Finally, in Fig.~\ref{cv_h_2} we compare the TMRG result with FTLM calculation on the chain $L=16$ 
with periodic boundary conditions at $T/J=0.5$ in order to establish a reliable comparison between
them. The two curves are in good agreement, especially in the vicinity of the two critical fields, with
some deviations in the center of the intermediate phase that are due to finite--size effects of FTLM data.

\section{Thermal Transport}
\label{sec:transport}
In this section we turn our attention to the transport properties of the $S=1$ large--$D$ model 
\eqref{s1model}. Within the linear response theory, the heat current $\mathcal{J}_Q$ and the spin 
current $\mathcal{J}_S$ are related to gradients of magnetic field $\nabla H$ and temperature 
$\nabla T$ by the transport coefficients $C_{ij}$ [\onlinecite{mahan}] :
\begin{equation}
\begin{pmatrix} \mathcal{J}_Q \\ \mathcal{J}_S \end{pmatrix} = 
\begin{pmatrix} C_{QQ} & C_{QS} \\ C_{SQ} & C_{SS} \end{pmatrix}
\begin{pmatrix} -\nabla T \\ \nabla H \end{pmatrix}\,,\nonumber
\end{equation}
where $C_{QQ}=\kappa_{QQ}$ ($C_{SS}=\sigma_{SS}$) is the heat (spin) conductivity. The coefficients 
$C_{ij}$ correspond to the dc limit of the real part of the appropriate current--current correlation 
functions (frequency--dependent conductivities), $C_{ij}=C_{ij}(\omega\to0)$. Note that under the 
assumption of vanishing spin current, which is relevant to certain experimental 
setups, the thermal conductivity $\kappa$ is redefined as follows:
\begin{equation}
\kappa=\kappa_{QQ}-\beta C_{QS}^2/C_{SS}\,,
\label{kmagc}
\end{equation}
where the second term is usually called the magnetothermal correction. Such a term originates from 
the coupling of the heat and spin currents in the presence of magnetic 
field [\onlinecite{louis},\onlinecite{sakai2005},\onlinecite{meisner}]. Here we present results for the heat conductivity 
$\kappa_{QQ}(\omega)$ calculated for $S=1$ model with FTLM on the chain up to $L=16$ sites and 
exact results obtained for $S=1/2$ model. In the latter case, we comment also on the
$\beta C_{QS}^2/C_{SS}$ term.

The real part of a given current--current correlation function (real 
part of the conductivity) can be written as:
\begin{equation}
C_{ij}(\omega)=2\pi D_{ij}\delta(\omega)+C^{\text{reg}}_{ij}(\omega)\,,
\label{cond}
\end{equation}
where the regular part $C^{\text{reg}}_{ij}(\omega)$ can be expressed in terms of eigenstates 
$|n\rangle$ and eigenenergies $\epsilon_n$:
\begin{eqnarray}
C^{\text{reg}}_{ij}(\omega)=\frac{\pi\beta^r}{L}\frac{1-e^{-\beta\omega}}{\omega} \nonumber\\
\sum_{\epsilon_n\ne\epsilon_m}p_n\langle m|\mathcal{J}_i|n\rangle
\times \langle n|\mathcal{J}_j|m\rangle\delta(\epsilon_n-\epsilon_m-\omega)\,,
\label{condreg}
\end{eqnarray}
while the dissipationless component with the Drude weight is related to the degenerate matrix 
elements:
\begin{equation}
D_{ij}=\frac{\beta^{r+1}}{2L}\sum_{\epsilon_n=\epsilon_m}p_n
\langle m|\mathcal{J}_i|n\rangle \langle n|\mathcal{J}_j|m\rangle\,,
\label{conddru}
\end{equation}
where $p_n=\exp(-\beta\epsilon_n)/Z$ are corresponding Boltzmann weights and $Z$ is the partition 
function.

In the case of heat conductivity, $C_{QQ}(\omega)=\kappa_{QQ}(\omega)$, $i=j=Q$, and $r=1$. The heat 
current $\mathcal{J}_Q=\sum_{n}j_{n}^{Q}$ can be defined by the lattice continuity equation 
$j^Q_{n}-j^Q_{n-1}=-\imath[{\mathcal H},{\mathcal H}_{n-1}]$, where ${\mathcal H}_n$ is the local 
energy density of \eqref{s1model}, with ${\mathcal H}=\sum_{n}{\mathcal H}_{n}$. Such a definition 
leads to
\begin{equation}
\mathcal{J}_Q=\sum_{n}\Big[J^2\mathbf{S}_{n-1}\cdot\Big(\mathbf{S}_{n}\times\mathbf{S}_{n+1}\Big)
+\left(2DS^{z}_{n}+H\right)j^{S}_{n}\Big]\,,
\end{equation}
where $j^S_n=J\left(S^{x}_{n}S^{y}_{n+1}-S^{y}_{n}S^{x}_{n+1}\right)$ is the local spin current. 
Note that in the presence of a finite magnetic field, $H\ne0$, the heat current $\mathcal{J}_Q$ is 
not simply equal to energy current $\mathcal{J}_{E}$ but instead is [\onlinecite{mahan}]
\begin{equation}
\mathcal{J}_Q=\mathcal{J}_{E}+H\mathcal{J}_S\,,
\label{cqths}
\end{equation}
with $\mathcal{J}_S=\sum_{n}j^{S}_{n}$.

\begin{figure}[!ht]
\includegraphics[width=\columnwidth]{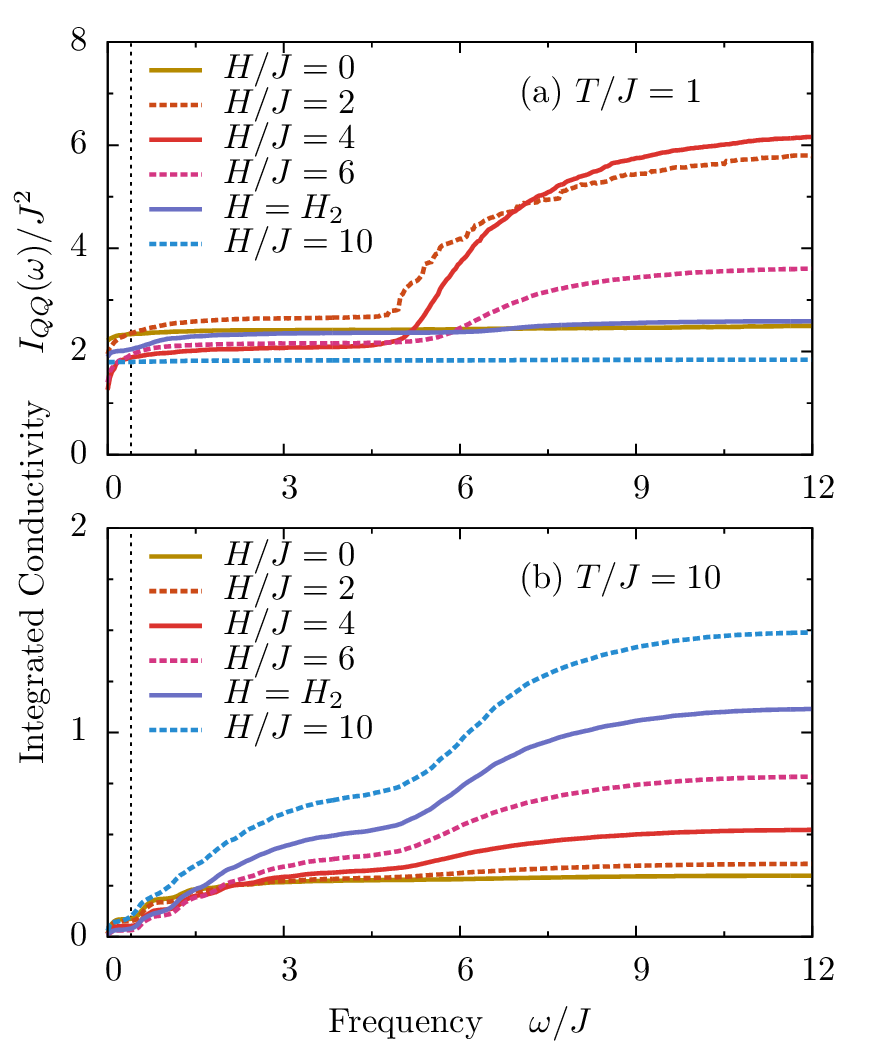}
\caption{(Color online) Integrated conductivity $I_{QQ}(\omega)$ for (a) $T/J=1$ and (b) $T/J=10$ 
as calculated for $L=16$ sites and different fields $H$. Dashed vertical line represents 
$\omega_0/J=2\pi/L\sim 0.4$.}
\label{integrated}
\end{figure}

Since our numerical calculation is performed on a finite chain, 
it is expected that the $\kappa_{QQ}(\omega)$ is a sum of weighted $\delta$ functions.
Therefore in Fig.~\ref{integrated} we present the integrated conductivity
\begin{equation}
I_{QQ}(\omega)=\frac{1}{2\pi}\int\limits_{-\omega}^{\omega}
\mathrm{d}\omega^\prime\,\kappa_{QQ}(\omega^\prime)\,,
\end{equation}
which is a much more reliable, monotonically increasing function, when numerically dealing with 
finite--system results.

From Fig.~\ref{integrated} it becomes apparent that $\kappa_{QQ}(\omega)$ exhibits two, well 
separated regions: the low--$\omega$ part and the high--$\omega$ part that is activated around 
$\omega/J\gtrsim D$. The spectral representation of $\kappa_{QQ}(\omega)$ of Eq.~\eqref{condreg} 
implies that nonzero matrix elements exist only for states $|n\rangle$ and $|m\rangle$ which obey 
the $\Delta S^z=0$ and $\Delta k=0$ selection rules. At low enough $T$, the high--frequency part 
of $\kappa^{\text{reg}}_{QQ}(\omega)$ should be dominated by transitions between the ground state 
and the next in energy state with the same total magnetization. As mentioned already, for $H<H_1$, 
the ground state $|\Omega\rangle$ carries zero azimuthal spin $S^z=0$ and the elementary 
excitations are the degenerate $S^z=1$ excitons and $S^z=-1$ antiexcitons with energy momentum 
dispersion $\epsilon(k)$ [\onlinecite{papanicolaou}] .The next in energy state that belongs to the total 
$S^z=0$ subspace is constructed by an exciton with crystal momentum $k_1$ and an antiexciton with 
$k_2$ and energy equal to $\epsilon(k_1)+\epsilon(k_2)$, which will be referred to as an 
exciton--antiexciton continuum. Therefore, at low $T$, the simplest possibility is a transition 
between the ground state and the exciton--antiexciton continuum at $k=k_1+k_2=0$, resulting 
contributions from a band of frequencies with boundary lines $\omega_{\alpha,\beta}$, where
\begin{equation}
\omega_{\alpha,\beta}=2D \mp 4 J +2 J^2/D\pm J^3/D^2\,.
\end{equation}

\begin{figure}[!ht]
\includegraphics[width=\columnwidth]{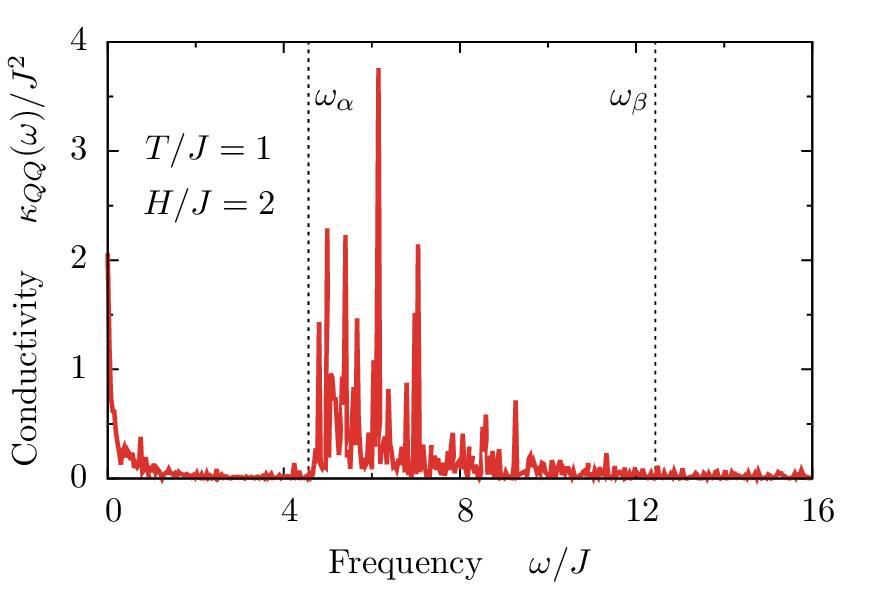}
\caption{(Color online) Frequency dependence of $\kappa_{QQ}(\omega)$ at $H=2$ and $T/J=1$. Labels 
$\omega_{\alpha,\beta}$ indicate the boundaries of the band with nonvanishing weight at low $T$.}
\label{kappa}
\end{figure}

In Fig.~\ref{kappa} we plot the frequency dependence of $\kappa_{QQ}(\omega)$ at $H=2$ and 
relatively low temperature $T/J=1$. As predicted, the high--frequency part of
$\kappa^{\text{reg}}_{QQ}(\omega)$ is activated at frequencies around $\omega_{\alpha}$ and
terminates at $\omega_{\beta}$, a result consistent with the preceding analysis.

For $H>H_2$ the ground state is fully polarized with no other state sharing the same $S^z$ 
subspace; therefore it is expected that contributions at high frequencies will vanish. This is 
supported by our numerical results and is evident in Fig.~\ref{integrated}(a), where for $H\geq 
H_{2}$ only the $\omega\sim0$ contributions are present. In the intermediate phase for 
$H_{1}<H<H_{2}$, the elementary excitations are difficult to calculate and there can be no 
analytical predictions such as lines $\omega_{\alpha,\beta}$. From the numerical data presented in 
Fig.~\ref{integrated}(a), we conclude that for $H_1<H<H_2$ the high--$\omega$ part of 
$\kappa^{\text{reg}}_{QQ}(\omega)$ is active at a band roughly between lines $\omega_{\alpha}$ and 
$\omega_{\beta}$ with intensity that is gradually reduced as $H\to H_{2}$.

Several conclusions can be drawn also for $\omega\to 0$ behavior of $\kappa_{QQ}(\omega)$. To begin 
with, in Fig.~\ref{integrated}(b) an anticipated result for nonintegrable systems is illustrated, 
namely, that Drude weight $D_{QQ}$ vanishes for high temperatures. On the other hand, at low 
temperatures, $D_{QQ}$ remains finite at any value of $H$, as can be seen in 
Fig.~\ref{integrated}(a). Moreover, for $H\geq J$ the $\omega\sim0$ contributions are dominant in the 
total sum rule $I_{QQ}(\omega=\infty)$ and almost all weight is in Drude weight itself. Since the 
model \eqref{s1model} is a nonintegrable, one would expect that $D_{QQ}$ is vanishing exponentially 
fast (at least for $T\to\infty$) with system size $L$, leading to diffusive transport in the 
thermodynamic limit [\onlinecite{haldane},\onlinecite{karadamoglou2004}].

\begin{figure}[!ht]
\includegraphics[width=\columnwidth]{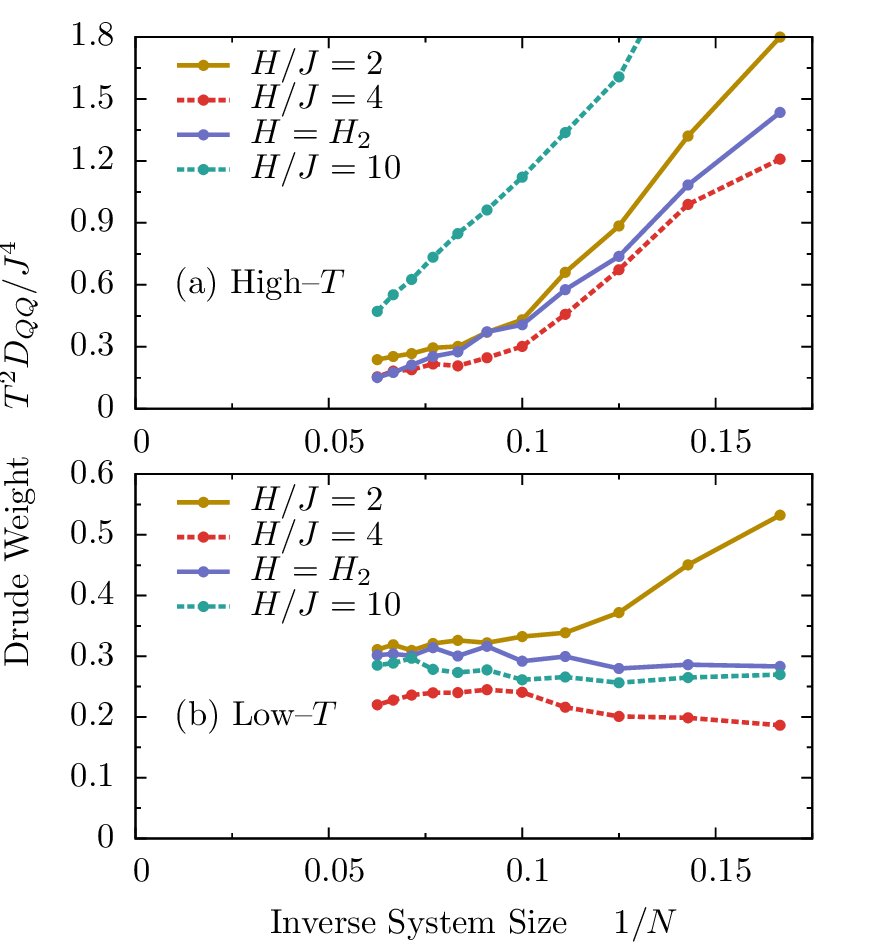}
\caption{(Color online) System size scaling of Drude weight $D_{QQ}$ at (a) $T/J=10$ and (b) 
$T/J=1$, obtained for systems with $L=6,\dots,16$ sites with various magnetic fields $H/J=2,4,8,10$.}
\label{size}
\end{figure}

In order to clarify this, we present in Fig.~\ref{size} inverse system size $1/L$ scaling of the 
$D_{QQ}$ for various values of $T$ and $H$. For $T\gg J$ the Drude weight is indeed vanishing 
exponentially fast, consistent with diffusive transport. However, this is not the case for 
low $T$, where the scaling of $D_{QQ}$ seems to weakly depend on system size. The choice of 
$H$ that determines whether the system is in the gapped or gapless phase does not seem to affect 
this scaling. Yet, a finite value of $D_{QQ}$ in the 
thermodynamic limit is one of the features of integrable systems [\onlinecite{zotos1997}], which is clearly 
not the case of the considered model \eqref{s1model} [\onlinecite{haldane},\onlinecite{karadamoglou2004}]. One of the 
possible explanations of this phenomenon is that the intrinsic diffusive processes at low $T$, 
that will result in a zero $D_{QQ}$ in the thermodynamic limit, become effective beyond the reachable 
system size or the energy resolution of the method presented here. As a result, it is expected that 
as one increases the system size, the spectral weight from $D_{QQ}$ shifts to 
$\kappa^{\text{reg}}_{QQ}(\omega<\omega_0)$, with $\omega_0/J\sim 2\pi/L$ [\onlinecite{long2003},\onlinecite{naef1998}].
The latter completely dominates the low--$\omega$ 
behavior of $\kappa_{QQ}(\omega)$ in the thermodynamic limit ($L\to\infty$). Therefore, to capture
this finite--size effect, in the following we will consider integrated conductivity
$I_{QQ}(\omega_0)$ (requency $\omega_0$ is depicted as vertical dashed line in Fig.~\ref{integrated}).

To gain insight into the origin of the slowly decaying Drude weight at low $T$, let us consider 
thermal transport in the effective low--energy $S=1/2$ Hamiltonian \eqref{s12model}. The heat 
current $\wt{\mathcal{J}}_Q$ is defined for this model in the same way, i.e., 
$\wt{j}^Q_{i}-\wt{j}^Q_{i-1}=-\imath[\wt{{\mathcal H}},\wt{{\mathcal H}}_{i-1}]$ with 
$\wt{{\mathcal H}}=\sum_{i}\wt{{\mathcal H}}_i$, leading to
\begin{equation}
\wt{\mathcal{J}}_Q=\sum_{n}\Big[4J^2\wt{\mathbf{S}}_{n-1}\cdot\Big(\wt{\mathbf{S}}_{n}\times
\wt{\mathbf{S}}_{n+1}^{\prime}\Big)+\wt{H}\wt{j}_{n}^{S}\Big]\,,
\end{equation}
with $\wt{\mathbf{S}}_{n}^{\prime}=(\wt{S}_{n}^{x},\wt{S}_{n}^{y},\Delta\wt{S}_{n}^{z})$. Other 
definitions and properties of the currents and conductivity remain the same [Eq.~\eqref{kmagc}-\eqref{conddru},\eqref{cqths}] 
with appropriate $\wt{\mathcal{J}}_\alpha$,
$\alpha=Q,E,S$ and $\wt{J}=2J$. 

It is known that the $S=1/2$ Heisenberg model is integrable, with heat current being one of the 
conserved quantities, $[\wt{\mathcal{J}}_Q,\wt{{\mathcal H}}]=0$, leading directly to its 
nondecaying behavior and within the linear response to infinite thermal conductivity. Also, the 
integrability of the model \eqref{s12model} makes the calculation of $\wt{D}_{QQ}$ feasible in the 
thermodynamic limit. As a consequence of Eq.~\eqref{cqths}, one can decompose Drude weight in 
terms of the energy and spin contribution
\begin{equation}
\wt{D}_{QQ}=\wt{D}_{EE}+2\beta\wt{H}\wt{D}_{ES}+\beta\wt{H}^{2}\wt{D}_{SS}\,,
\end{equation}
where Drude weights are defined in Eq.~\eqref{conddru}, with $r=1$ for $i=j=Q$ or $i=j=E$, and 
$r=0$ for $i=j=S$ or $i=E,\,j=S$.

The $\wt{D}_{EE}$ and $\wt{D}_{ES}$ at finite temperatures have been calculated by Sakai and 
Kl\"{u}mper [\onlinecite{sakai2005}] using a lattice path integral formulation, where a quantum transfer 
matrix (QTM) in the imaginary time is introduced. Correlations and thermodynamic quantities can be 
evaluated in terms of the largest eigenvalue of the QTM. The importance of this method yields to 
the fact that all quantities are found by solving two nonlinear integral equations at 
arbitrary magnetic fields, temperatures and anisotropy parameters. Here we repeat the calculation 
using $\Delta=1/2$.

On the other hand, spin Drude weight $\wt{D}_{SS}$ at finite magnetic field is computed based on a 
generalization of a method that was proposed by Zotos [\onlinecite{zotos1999}], where $\wt{D}_{SS}$ was 
calculated using the Bethe ansatz technique at zero magnetic field. The presence of magnetic field 
will cause some changes to the TBA equations [\onlinecite{takahashi}], but the overall analysis is 
essentially the same. 

\begin{figure}[!ht]
\includegraphics[width=\columnwidth]{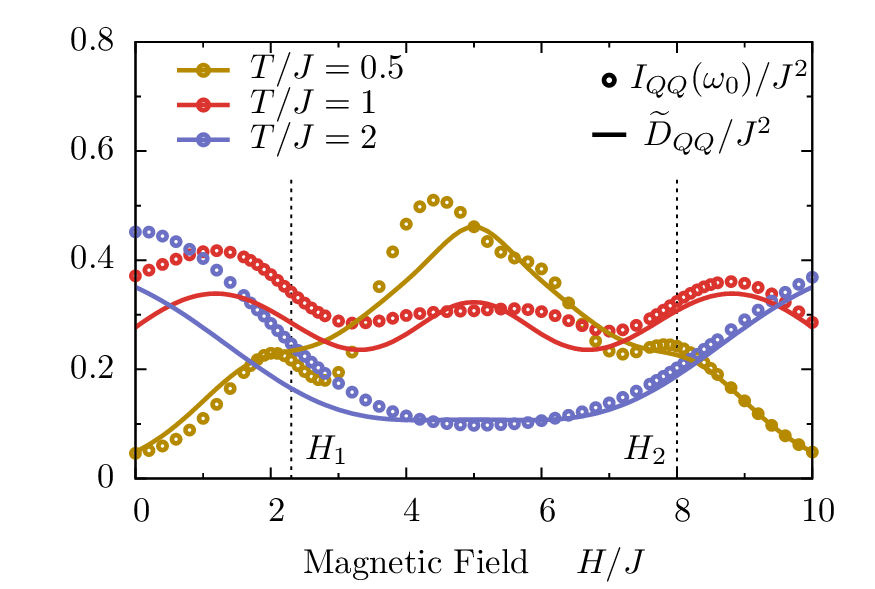}
\caption{(Color online) Comparison of $S=1$ integrated conductivity $I_{QQ}(\omega_0)$ at 
$\omega_0=2\pi/L$ for $L=16$ with exact $S=1/2$ Drude weight $\wt{D}_{QQ}$ calculated in the 
thermodynamic limit for $T=0.5,1$ and $2$ as a function of the magnetic field $H$.}
\label{transport}
\end{figure}

In Fig.~\ref{transport} we compare $\wt{D}_{QQ}$ for the $S=1/2$ model with the numerically obtained 
integrated conductivity $I_{QQ}$ at $\omega_0$ for the $S=1$ model on $L=16$ sites. As is clearly 
visible, the overall agreement is satisfactory. The magnetic field dependence of Drude weight 
$\wt{D}_{QQ}$ includes all characteristic features of the $S=1$ low--$\omega$ behavior. From the 
results obtained for the thermal transport, as in the case of magnetization and specific heat, we 
observe that the mapping is much more accurate close to $H_{2}$ than close to $H_{1}$. Also, due to 
spin--inversion symmetry, the $S=1/2$ results are symmetric with respect to $H=5$ ($\wt{H}=0$), 
where lack of such a symmetry for the $S=1$ model is expected.

\begin{figure}[!ht]
\includegraphics[width=\columnwidth]{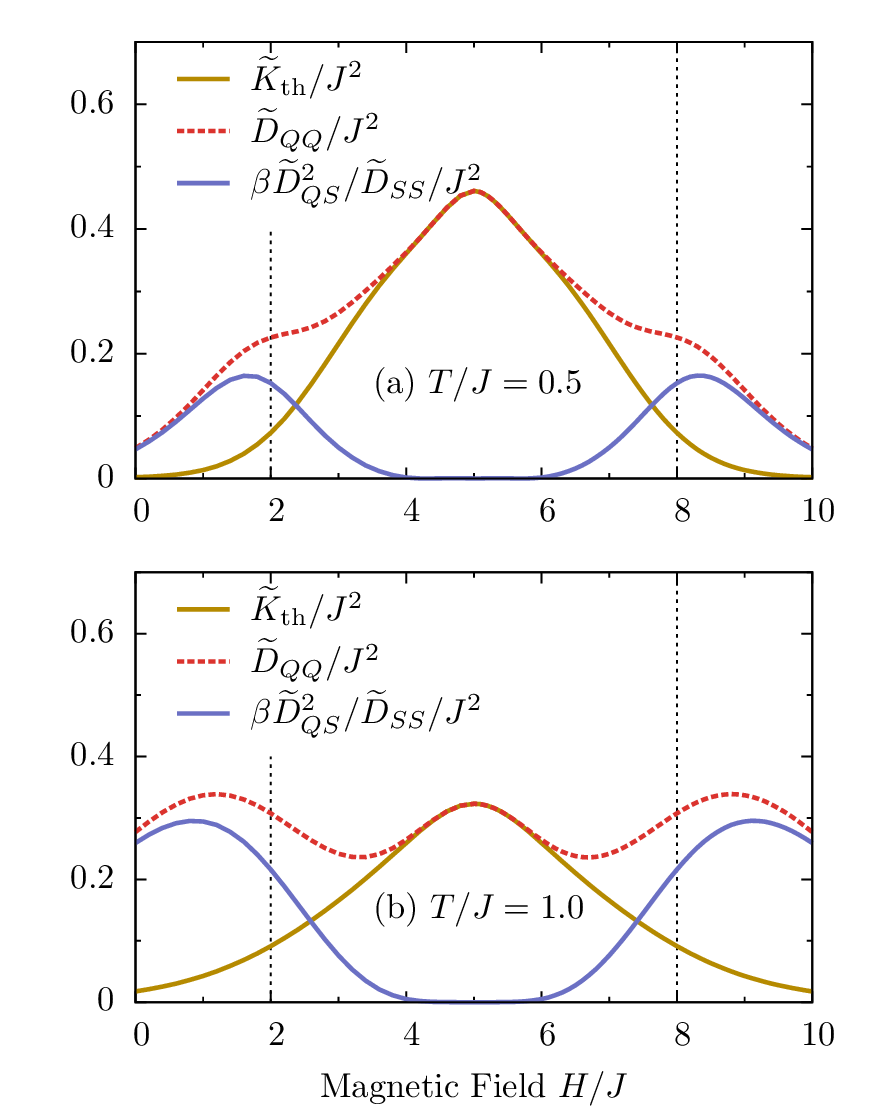}
\caption{(Color online) Magnetic field dependence of $\wt{D}_{QQ}$, $\wt{K}_{\text{th}}$ and MTC 
term at fixed temperature (a) $T/J=0.5$ and (b) $T/J=1$. Vertical lines indicate the critical 
fields.}
\label{drude_s12}
\end{figure}

Let us now comment on the magnetothermal corrections (MTC) to heat conductivity [Eq.~\eqref{kmagc}] 
for the $S=1/2$ model. Frequency--dependent thermal conductivity $\kappa$ can be written in the same form 
as Eq.~\eqref{cond}, with the weight of the singular part given by [\onlinecite{mahan}]
\begin{equation}
\wt{K}_{\text{th}}=\wt{D}_{QQ} -\beta\wt{D}_{QS}^2/\wt{D}_{SS}\,,
\label{kth}
\end{equation}
where $r=0$ for $i=Q,\,j=S$. Both of the two competing terms that contribute to 
$\wt{K}_{\text{th}}$ become important at finite magnetic fields. In Fig.~\ref{drude_s12} we depict 
the magnetic field dependence of $\wt{D}_{QQ}$, $\wt{K}_{\text{th}}$, and the MTC term at fixed 
temperature (a) $T/J=0.5$ and (b) $T/J=1$, as have been calculated for the $S=1/2$ model 
\eqref{s12model}.

As expected, the MTC term is exactly zero at the zone center ($\wt{H}=0$) but it becomes finite at 
finite $H$, where we see a bell curve behavior, with the peak centered close to the critical fields 
at low $T$. Upon increasing $T$, the position of the first (second) peak is shifted to lower 
(higher) magnetic fields. While $\wt{D}_{QQ}$ exhibits a pronounced nonmonotonic behavior as a 
function of $H$, with two peaks centered close to the critical fields, the inclusion of the second 
term of Eq.~\eqref{kth} results in an overall suppression of $\wt{K}_{\text{th}}$ and the 
cancellation of this behavior. This finding is confirmed by a numerical study of the thermal 
transport in the $S=1/2$ \textit{XXZ} chain in the presence of a magnetic field [\onlinecite{meisner}] based on exact
diagonalization of a finite chain.

In all cases considered here, the thermal conductivity at $T<J$ has a maximum located at
$H\simeq H_m=(H_1+H_2)/2$. However, this is not what is observed in the experiment. The thermal conductivity 
measurements at low $T$ of the DTN compound[\onlinecite{Kohama},\onlinecite{Sun}] exhibit sharp peaks in the vicinity of critical
fields $H_{1,2}$. Detailed analysis of spin contribution to the total thermal conductivity
is a nontrivial task due to the presence of phononic contribution. Also, the DTN compound is a quasi--1D material
with $J_\perp/J\simeq0.18$, and for temperatures below $T_{N}<1.2\,\mathrm{K}$ ($T/J\lesssim 0.5$)
is in a 3D ordered state [\onlinecite{zapf},\onlinecite{yin},\onlinecite{Zvyagin1},\onlinecite{Mukhopadhyay}] with long--range correlations [\onlinecite{Chiatti},\onlinecite{Mukhopadhyay}].

\section{Electron Spin Resonance}
\label{sec:esr}

Electron spin resonance has been one of the main tools for experimental investigation of DTN [\onlinecite{Zvyagin3}] 
for a wide field range including the intermediate region $H_1<H<H_2$. 
The original experiment was repeated in Ref.~[\onlinecite{psaroudaki}] in order to clarify 
certain important features predicted by theory [\onlinecite{papanicolaou1}] such as the occurrence of a two--magnon bound state 
for strong fields in the region $H>H_2$. One of the main conclusions of the above references is that the essential features 
of the ESR spectrum observed in DTN are accounted for by the strictly 1D $S=1$ model \eqref{s1model}. 
Yet, even within this 1D model, calculation of the ESR spectrum has been difficult especially for fields in the 
intermediate phase. 

It is the purpose of the present section to investigate the structure of the zero--temperature low--lying ESR spectrum throughout the 
intermediate region $H_1<H<H_2$ using the mapping to the effective $S=1/2$ model \eqref{s12model} for which a 
rigorous solution can be obtained using the Bethe ansatz. As a preparation for our main result, we recall that the 
extent of the intermediate phase predicted by the $S=1/2$ \textit{XXZ} model is given by $-\wt{H}_c<\wt{H}<\wt{H}_c$, where 
$\wt{H}_c=2 J(1+\Delta)=3J$ for $\Delta=1/2$. Upon translating this prediction in terms of the original field $H=\wt{H}+J+D$, 
the extent of the intermediate phase is given by
\begin{equation}
H_1=D-2 J\,, \qquad H_2=D+4 J\,, 
\label{ESRfields}
\end{equation}
where $H_2$ coincides with the exact upper critical field of Eq.~\eqref{H2} predicted by the $S=1$ model, whereas $H_1$ is an 
approximate prediction for the lower critical field that is consistent with Eq.~\eqref{H1}, restricted to first order in 
the $1/D$ expansion. Accordingly, the field dependence of the ESR spectrum outside the intermediate phase is given by
\begin{eqnarray}
\omega_B=D+2 J-H  \quad \text{for} \quad H<H_1\,,\nonumber \\
\omega_C=H-D  \quad \text{for} \quad H>H_2\,, 
\label{omegaBC}
\end{eqnarray}
where $\omega_C$ is the $k=0$ value of the magnon dispersion for $\wt{H}>\wt{H}_c$, and $\omega_B$ is the corresponding 
value for $\wt{H}<-\wt{H}_c$. 
Note that $\omega_C$ coincides with the exact value of the corresponding prediction in the $S=1$ model, whereas $\omega_B$ 
is again the first order approximation within a systematic $1/D$ expansion [\onlinecite{psaroudaki}].

The preceding elementary calculation of the ESR spectrum cannot be simply extended into the intermediate phase even within 
the effective $S=1/2$ model. However, recent developments in the Bethe ansatz method [\onlinecite{kitanine},\onlinecite{caux}] allow 
the semi analytical evaluation of matrix elements between eigenstates in the $S=1/2$ Heisenberg model for any 
magnetization: the calculations reduce to the numerical evaluation of determinants of the order of 
the size of the spin system. When applied to the ESR operator
$|\langle m|\wt{S}_{\text{tot}}^{-}|\wt{\Omega}\rangle|^2$, where $|\wt{\Omega}\rangle $ is the 
ground state, $|m\rangle$ an excited state and $\wt{S}_{\text{tot}}^{-}=\sum_n \wt{S}_n^-$, it is 
found that there is essentially only one excited state, $|m^*\rangle$, that has significant weight 
in the spectrum. This state is a highly unusual one in the Bethe ansatz literature. While usually 
eigenstates are characterized by sets of real pseudomomenta $\lambda$ or pseudomomenta with 
imaginary parts symmetrically arranged around the real axis (``strings''), this state has all the 
$\lambda$'s real except one that is complex with an imaginary part $\imath\pi/2$. The existence of this 
state was recently discussed  [\onlinecite{ovch}] and it physically corresponds to a uniform change of the 
$\wt{S}^z$ component of the magnetization by 1. It is fascinating that the ESR experiments exactly probe this state and its dynamics.

From a computational point of view, it turns out to be rather difficult to find the pseudomomenta 
$\lambda$ for this state. The nonlinear Bethe ansatz equations at finite magnetization, in general, 
do not converge by iteration. To circumvent this problem, it was suggested  [\onlinecite{baxter}] to study 
chains with an odd number $N$ of spins, where indeed the problem is far less crucial  [\onlinecite{caux}]. In 
the following we present data for the magnetic field $\wt{H}$ dependence of the ESR resonance 
frequency $\omega_{m^*}=\epsilon_{m^*}-\epsilon_{\wt{\Omega}}$ and of the ESR matrix element 
$|\langle m^*|\wt{S}_{\text{tot}}^{-}|\wt{\Omega}\rangle|^2$ for $N=51$.The quantum numbers 
characterizing the ground state $|\wt{\Omega}\rangle$ with $M$ reversed spins are given by 
$I_{j=1,M}=-M/2+1,\dots,+M/2$, corresponding to a magnetization $\wt{S}^z=N/2-M$. The excited state 
$|m^*\rangle$ has $M+1$ reversed spins and is characterized by the quantum numbers 
$I_{j=1,M}=-M/2+1/2,\dots,+M/2-1/2$, $I_M=(N+M)/2$. 

\begin{figure}[!ht]
\includegraphics[width=\columnwidth]{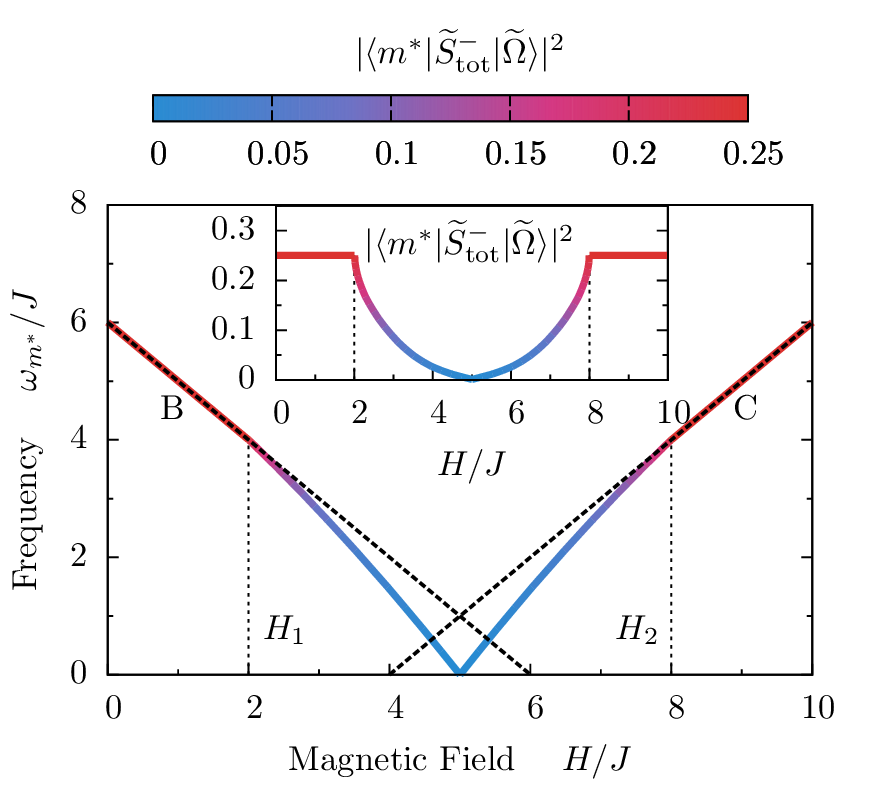}
\caption{(Color online) Field dependence of $T=0$ low--lying ESR lines calculated from the effective $S=1/2$ model diagonalized 
through the Bethe ansatz. Lines $B$ and $C$ are the straight lines $\omega_B$ and $\omega_C$ given in Eq.~\eqref{omegaBC} 
for fields outside the intermediate phase but bend downwards in a nontrivial manner upon entering the intermediate phase 
to meet at the center and thus form a $V$--like structure. The inset depicts the field dependence of the matrix element 
$|\langle m^*|\wt{S}_{\text{tot}}^{-}|\wt{\Omega}\rangle|^2$, which is directly relevant for the calculation of the intensity 
of ESR modes. Vertical dotted lines indicate the location of the critical fields $H_1$ and $H_2$ calculated from 
Eq.~\eqref{ESRfields}. }
\label{esr}
\end{figure}

The results of this intriguing calculation are summarized in Fig.~\ref{esr}, which depicts the field dependence of the
low--lying ESR lines as a function of the field $H$. As expected, these coincide with the straight lines $\omega_B$ and 
$\omega_C$ of Eq.~\eqref{omegaBC} for fields $H$ outside the intermediate phase, which bend downwards upon entering 
the intermediate phase to meet at the center and thus form a $V$--like structure. The calculated slope is $\pm3/2$ 
at the center and $\pm1$ at and beyond the edges of the intermediate phase. Also shown in Fig.~\ref{esr} is the calculated 
field dependence of the matrix element $|\langle m^*|\wt{S}_{\text{tot}}^{-}|\wt{\Omega}\rangle|^2$, which 
vanishes at the center but reaches a finite value $1/4$ that remains constant for all fields outside the intermediate 
phase. 

The currently predicted $V$--like ESR spectrum with vanishing intensity at its center is consistent with our earlier 
prediction  [\onlinecite{psaroudaki}] made by a rough numerical calculation on small ($N=10$) chains within the $S=1$ model \eqref{s1model},
but disagrees with a $Y$--like structure with nonvanishing intensity at the center made 
by Cox {\it et al.}  [\onlinecite{cox}] by a calculation within the same $S=1$ model. Concerning possible experimental observation, 
the rapid vanishing of intensity near the center would make the $V$--mode especially sensitive to small perturbations that 
are ever present in effective Heisenberg models  [\onlinecite{Zvyagin3},\onlinecite{psaroudaki}].

Some caution is necessary with regard to the results presented in this section concerning the structure of the ESR spectrum in the intermediate phase. As stated earlier, most of the intensity is concentrated on a single resonance frequency $\omega_{m^*}$ with a $\delta$--function line shape, emerging from transitions between the ground state and the excited state $|m^*\rangle$. Apart from this dominant contribution, the Bethe ansatz calculation revealed that the ESR spectrum consists of secondary transitions with small, but non vanishing intensity. These transitions correspond to resonance frequencies that lie above $\omega_{m^*}$ with negligible matrix elements and are thus omitted from Fig.~\ref{esr}. These secondary peaks exist throughout the intermediate phase for $-\wt{H}_{c}<\wt{H}<\wt{H}_c$ but lose their intensity for $\wt{H} \geq \wt{H}_c$ and $\wt{H} \leq -\wt{H}_c$. In this case, the only ESR transition is the one between the ferromagnetic ground state and the $k=0$ single magnon, with resonance frequency
\begin{eqnarray}
\omega_{sm}&=&2 J(1-\Delta) +\wt{H}  \quad \text{for} \quad \wt{H}\geq \wt{H}_c\,,\nonumber \\
&=&2 J(1-\Delta) -\wt{H}  \quad \text{for} \quad \wt{H}\leq -\wt{H}_c\,.
\label{omegasm}
\end{eqnarray}

In order to clarify this more complicated ESR spectrum, two limiting cases are considered; the isotropic chain ($\Delta=1$) and the \textit{XY} model ($\Delta=0$).  In the presence of isotropic interaction, the resonance frequency $\omega_{sm}=\wt{H}$ with a $\delta$--function line shape is extended in the intermediate region. The line is precisely at the Zeeman energy for any magnetic field, with intensity that gradually vanishes as $\wt{H}\rightarrow 0$. In the presence of a small perturbation to the isotropic Hamiltonian, the ESR spectrum is again dominated by a single line, but the presence of anisotropy causes a shift in the position of the resonance peak that varies with magnetic field  [\onlinecite{AffleckESR}].

On the other hand, the picture gets more involved for $\Delta=0$. A numerical calculation performed by Maeda and Oshikawa  [\onlinecite{maeda}] showed that the single magnon picture with a $\delta$--function line shape at $\omega_{sm}=2J\pm \wt{H}$ holds only for $\wt{H}\geq \wt{H}_c$ and $\wt{H}\leq -\wt{H}_c$. This picture breaks down in the intermediate phase, where absorption takes place over a finite frequency range with boundaries $2\wt{H}<\omega<4J$.

From the discussion above it follows that the value of anisotropy considered here, $\Delta=1/2$, lies approximately in the middle of the $0\leq\Delta\leq1$ region, combining features from both extreme cases. The argument of a single line is substantially correct and adequately describes the ESR spectrum, while secondary peaks exist with negligible intensity. These peaks will evolve into a band of resonance frequencies in the $\Delta=0$ limit.

\section{Conclusions}
\label{sec:conclusions}
We have investigated the thermodynamic and dynamical properties of the one--dimensional $S=1$ antiferromagnetic chain 
with large easy plane anisotropy, in the presence of a uniform magnetic field. 
An effective $S=1/2$ Heisenberg \textit{XXZ} Hamiltonian is derived based on a mapping of the original $S=1$ Hamiltonian 
into its low--energy subspace, which enable us to gain a better physical understanding of the considered model. For all 
quantities studied here, results for both the $S=1$ and $S=1/2$ model are presented and compared in order to test the 
effectiveness of the mapping, and results from the exactly solvable \textit{XXZ} model are collated to complete the theoretical 
description. 

The temperature and magnetic field dependence of magnetization and specific heat of the $S=1$ model 
have been studied using a TMRG algorithm, which allows us to obtain these quantities with satisfactory accuracy in 
the thermodynamic limit. The thermodynamic Bethe ansatz is applied to derive the same quantities for the $S=1/2$ model.  
The critical exponent that describes the behavior of magnetization near the critical fields at very low $T$ 
is extracted from the numerical data of the $S=1$ model and found equal to $\delta=2$. 
This result renders the considered model in the 
same universality class as a broad collection of various models of quantum magnetism. 
Furthermore, the temperature dependence of magnetization for both models reveals the existence of extrema at 
some temperature $T_c$, which is interpreted
as the critical temperature below which the description of the system in terms of Luttinger liquid is valid. 
A magnetic phase diagram is constructed that represents the crossover into a low--T Luttinger liquid regime. 
The section of thermodynamics is completed with the investigation of specific heat as a function of $H$ and $T$. The 
$\mathcal{C}_v(H)$ curve exhibits a characteristic double peak around critical fields $H_{1,2}$, 
and the $\mathcal{C}_v(T)$ curve reveals a linear dependence at low $T$, consistent with the LL phase. 

We also give a description of the heat conductivity $\kappa_{QQ}$, calculated for the $S=1$ model with a FTLM algorithm 
on a finite chain of length $L=16$. We observe that the singular part of $\kappa_{QQ}$, namely, the Drude peak $D_{QQ}$, 
vanishes for high $T$, an anticipated result for nonintegrable systems. On the contrary, at low $T$, $D_{QQ}$ remains the 
significant contribution to the total sum rule of $\kappa_{QQ}$ at all considered fields. Therefore the low-$\omega$ part of the integrated
conductivity $I_{QQ}$ is compared with the $S=1/2$ Drude weight $\wt{D}_{QQ}$ 
calculated in the thermodynamic limit. The overall agreement is satisfactory, with $\wt{D}_{QQ}$ including 
all the characteristic features of the $S=1$ behavior. Within the integrable $S=1/2$ model, the heat current 
$\mathcal{J}_Q$ is a conserved quantity giving infinite thermal conductivity. 
Nevertheless, it is a nontrivial question as to which extent integrability of the low--energy effective $S=1/2$ 
Hamiltonian influences transport properties of the full $S=1$ model. However, this is beyond the 
scope of this paper, and we leave it as a motivation for further studies.

Finally, the low--lying ESR spectrum of the effective $S=1/2$ model is analyzed for fields in the intermediate region 
in order to complete earlier work on the $S=1$ model. A semi analytical evaluation based on the Bethe ansatz predicts 
that ESR lines form a $V$--like structure in the low-lying intermediate phase with vanishing intensity at its center.   

Concerning the experimental observations of the results presented throughout the paper, we conclude that measurements on DTN 
showed that some characteristics expected for a one--dimensional system are not present, indicating that the system exhibits 3D behavior. 
In the case of thermal conductivity, not only the dimensionality of the system, 
but the inclusion of scattering mechanisms such as phonons are necessary in order to reach a realistic description.  

\acknowledgments
This work was supported by the European Commission through the LOTHERM Project (FP7-238475);
the European Union (European Social Fund, ESF), and Greek national funds through the Operational
Program ``Education and Lifelong Learning'' of the NSRF under ``Funding of proposals that have received
a positive evaluation in the 3rd and 4th call of ERC Grant Schemes''; the European Union Program
No. FP7-REGPOT-2012-2013-1 under Grant No.~316165; and the Slovenian Agency Grant No.~P1-0044.
\setcounter{equation}{0}
\appendix
\renewcommand{\theequation}{A\arabic{equation}}
\section{Effective Hamiltonian}
\label{sec:ap1}
Here we give more details about the derivation of the effective spin Hamiltonian. 
For $H<H_1$ the ground state $|\Omega\rangle$ and lowest excitations $|\Psi_1\rangle$ and 
$|\Psi_2\rangle$ are
\begin{equation}
|\Omega\rangle=|1,0\rangle \otimes |1,0\rangle \otimes |1,0\rangle \otimes |1,0\rangle
\otimes \dots \otimes |1,0\rangle\,,\nonumber
\end{equation}
\begin{equation}
|\Psi_{1,2}\rangle=\frac{1}{\sqrt{N}}\sum_n \mathrm{e}^{\imath k n}|n_{\mp}\rangle\,,
\label{Ap1}
\end{equation}
where states $|n_{-}\rangle$ and $|n_{+}\rangle$ carry nonzero azimuthal spin equal to $-1$ and 
$+1$ respectively only at the site $n$. At zero magnetic field the states $|\Psi_1\rangle$ and 
$|\Psi_2\rangle$ are degenerate with a known energy momentum dispersion 
$\epsilon(k)$  [\onlinecite{papanicolaou}]. This degeneracy is lifted at nonzero magnetic field $H$ due to 
the Zeeman energy. Upon increasing $H$ the state $|\Psi_1\rangle$ approaches the ground state, 
whereas the energy difference of states $|\Psi_1\rangle$ and $|\Psi_2\rangle$ equals $2 H$ and 
becomes larger. Close to $H_1$ the low--energy space is spanned only by states $|\Psi_1\rangle$ and 
$|\Omega\rangle$ and the contribution of $|\Psi_2\rangle$ can be neglected. A new $S=1/2$ 
representation can be used:
\begin{equation}
|\wt{\Omega}\rangle=|\downarrow\rangle \otimes |\downarrow\rangle \otimes |\downarrow\rangle
\otimes |\downarrow\rangle \otimes \dots \otimes |\downarrow\rangle\,, \nonumber
\end{equation}
\begin{equation}
|\wt{\Psi_1 }\rangle=\frac{1}{\sqrt{N}}\sum_n \mathrm{e}^{\imath k n}|\wt{n}\rangle\,,
\label{Ap2}
\end{equation}
where state $|\wt{n}\rangle$ differs from $|\wt{\Omega}\rangle$ by a spin--up at site $n$. 
Therefore, we project the original Hamiltonian \eqref{s1model} into this subspace, and the 
resulting effective Hamiltonian up to a constant is:
\begin{equation}
\wt{\mathcal{H}}=\sum_{n}\left[2J\left(\wt{S}_n^{x}\wt{S}_{n+1}^{x}+\wt{S}_n^{y}\wt{S}_{n+1}^{y}
+\Delta\wt{S}_n^{z}\wt{S}_{n+1}^{z}\right)+\wt{H}\wt{S}_n^{z}\right]\,,
\label{Ap3}
\end{equation}
where $\Delta=1/2$ and $\wt{H}=-J-D+H$. 

For $H>H_2$ the fully FM ground state and the single magnon eigenstate are:
\begin{equation}
|\Omega \rangle=|1,-1\rangle \otimes |1,-1\rangle \otimes |1,-1\rangle \otimes |1,-1\rangle
\otimes \dots \otimes |1,-1\rangle\,,\nonumber
\end{equation}
\begin{equation}
|\Psi\rangle=\frac{1}{\sqrt{N}}\sum_n \mathrm{e}^{\imath k n} | n \rangle\,,
\label{Ap4}
\end{equation}
where state $|n\rangle$ differs from the ground state by the fact that $S_n^{z}=0$. By identifying 
these two states with the $S=1/2$ states given in Eq.~\eqref{Ap2} the resulting model is again described by the 
Hamiltonian \eqref{Ap3}.

\renewcommand{\theequation}{B\arabic{equation}}
\section{Thermodynamic Bethe ansatz equations}
\label{sec:ap2}
According to the thermodynamic Bethe ansatz, a system of nonlinear integral equations provides all 
the required information for the calculation of the free energy of model \eqref{s12model} in the 
thermodynamic limit  [\onlinecite{takahashi}]. The number of these equations is determined by the value of 
parameter $\Delta$. For $\Delta=\cos(\pi/n)$ there are $n$ such equations with $f_{i}(x)$ unknown 
functions, where $i=1,2,\dots,n$. In the case we are studying here, we have $\Delta=1/2$ and $n=3$; therefore the full set of equations is
\begin{eqnarray}
\ln[1+f_1(x)]=-\frac{2 J}{T} 3 \sqrt{3}\,\delta(x)\,,\nonumber\\
\ln f_2(x)=-\frac{2 J}{T} 3 \sqrt{3}\, g(x)\nonumber\\
+\int\limits_{-\infty}^{\infty}\mathrm{d}y\,
g(x-y)\ln\left[1+2f_3(y)\cosh(3\wt{H}/2 T)+f_3(y)^2\right]\,,\nonumber\\
\ln f_3(x)=\int\limits_{-\infty}^{\infty}\mathrm{d}y\,g(x-y)\ln\left[1+f_2(y)\right]\,,\nonumber\\
\end{eqnarray}
where $g(x)=\mbox{sech}(\pi x/2)/4$. The above equations are solved numerically by an iterative 
process, where we generate a sequence of improving approximate solutions that converge rapidly. 
Once function $f_2(x)$ is determined, the free energy is given from
\begin{equation}
\wt{F}=\int\limits_{-\infty}^{\infty}\mathrm{d}x\,g(x)\ln[1+f_2(x)]\,.
\end{equation}

The specific heat and magnetization are given by
\begin{equation}
\wt{\mathcal{C}}_v=\beta^{2}\frac{\partial^2 \wt{F}}{\partial \beta^2}\,,\quad 
\wt{M}=-\frac{\partial \wt{F}}{\partial \wt{H}}\,,
\label{M}
\end{equation}
where $\beta=1/T$ is the inverse temperature. To avoid numerical differentiation, one can derive 
similar nonlinear equations and directly calculate the derivatives.



\begin{thebibliography}{99}
\bibitem{haldane} F. D. M. Haldane, Phys. Lett. A {\bf 93}, 464 (1983).
\bibitem{langari} A. Langari, F. Pollmann, and M. Siahatgar, J. Phys.: Condens. Matter {\bf 25}, 
406002 (2013).
\bibitem{papanicolaou} N. Papanicolaou, P.N. Spathis, J. Phys.: Condens. Matter {\bf 1}, 5555 (1989); 
Phys. Rev. B {\bf 52}, 16001 (1995).
\bibitem{hamer} A. F. Albuquerque, C. J. Hamer, and J. Oitmaa, Phys. Rev. B {\bf 79}, 054412 (2009).
\bibitem{papanicolaou1} N. Papanicolaou, A. Orend\'{a}\v{c}ov\'{a}, and M. Orend\'{a}\v{c}, Phys. 
Rev. B {\bf 56}, 8786 (1997).
\bibitem{zapf} V. S. Zapf, D. Zocco, B. R. Hansen, M. Jaime, N. Harrison, C. D. Batista, M. Kenzelmann, 
C. Niedermayer, A. Lacerda, and A. Paduan-Filho, Phys. Rev. Lett. {\bf 96}, 077204 (2006).
\bibitem{zapf2014} V. Zapf, M. Jaime, and C. D. Batista, Rev. Mod. Phys. {\bf 86}, 563 (2014).
\bibitem{yin} L. Yin, J.S. Xia, V. S. Zapf, N. S. Sullivan, and A. Paduan-Filho, Phys. Rev. Lett. 
{\bf 101}, 187205 (2008).
\bibitem{giamarchi} T. Giamarchi, and A. M. Tsvelik, Phys. Rev. B {\bf 59}, 11398 (1999); F. Mila, 
Eur. Phys. J. B {\bf 6}, 201 (1998).
\bibitem{psaroudaki} C. Psaroudaki, S. A. Zvyagin, J. Krzystek, A. Paduan-Filho, X. Zotos, and N. 
Papanicolaou, Phys. Rev. B {\bf 85}, 014412 (2012).
\bibitem{Zvyagin1} S. A. Zvyagin, J. Wosnitza, C. D. Batista, M. Tsukamoto, N. Kawashima, J. Krzystek, V. S. Zapf, M. Jaime, 
N. F. Oliveira Jr., and A. Paduan-Filho, Phys. Rev. Lett. {\bf 98}, 047205 (2007).
\bibitem{Zvyagin2} S. A. Zvyagin, C. D. Batista, J. Krzystek, V. S. Zapf, M. Jaime, A. Paduan-Filho, and 
J. Wosnitza, Physica B {\bf 403}, 1497 (2008).
\bibitem{Xiang} X. Wang, T. Xiang, Phys. Rev. B {\bf 56}, 5061 (1997); N. Shibata, J. Phys. Soc. 
Jpn. {\bf 66}, 2221 (1997).
\bibitem{prelovsek2013} For a recent review, see P.~Prelov\v{s}ek and J.~Bon\v{c}a, in {\it Strongly Correlated Systems - Numerical Methods}, edited by A.~Avella and F.~Mancini (Springer Series in Solid--State Sciences Vol. 176 (Springer,Berlin, 2013), pp. 1--29.
\bibitem{takahashi} M. Takahashi and M. Suzuki, Prog. Theor. Phys. {\bf 48}, 2187 (1972).
\bibitem{Affleck} I. Affleck, Phys. Rev. B {\bf 43}, 3215 (1991).
\bibitem{Chitra} R. Chitra, and T. Giamarchi, Phys. Rev. B {\bf 55}, 5816 (1997).
\bibitem{Sukai} T. Sakai, and M. Takahashi, Phys. Rev. B {\bf 57}, R8091 (1998).
\bibitem{Yang} C. N. Yang, and C. P. Yang, Phys. Rev. {\bf 150}, 327 (1966); Phys. Rev. {\bf 151}, 
258 (1966).
\bibitem{Wang} X. Wang, and L. Yu, Phys. Rev. Lett. {\bf 84}, 5399 (2000).
\bibitem{Wessel} S. Wessel, M. Olshanii, and S. Haas, Phys. Rev. Lett. {\bf 87}, 206407 (2001).
\bibitem{ruegg} C. R\"{u}egg, K. Kiefer, B. Thielemann, D. F. McMorrow, V. Zapf, B. Normand, M.B. 
Zvonarev, P. Bouillot, C. Kollath, T. Giamarchi, S. Capponi, D. Poilblanc, D. Biner, and K.W. 
Kr\"{a}mer, Phys. Rev. Lett. {\bf 101}, 247202 (2008).
\bibitem{sologubenko}A. V. Sologubenko, T. Lorenz, J. A. Mydosh, B. Thielemann, H. M. R{\o}nnow, 
C. R\"{u}egg, K. W. Kr\"{a}mer, Phys. Rev. B {\bf 80}, 220411(R) (2009).
\bibitem{Bouillot} P. Bouillot, C. Kollath, A.M. L\"{a}uchli, M. Zvonarev, B. Thielemann, C. 
R\"{u}egg, E. Orignac, R. Citro, M. Klanj\v{s}ek, C. Berthier, M. Horvati\'{c}, and T. Giamarchi, 
Phys. Rev. B {\bf 83}, 054407 (2011).
\bibitem{Maeda} Y. Maeda, C. Hotta, and M. Oshikawa, Phys. Rev. Lett. {\bf 99}, 057205 (2007).
\bibitem{Paduan} A. Paduan-Filho, X. Gratens, and N.F. Oliveira,Jr., Phys. Rev. B {\bf69}, 020405 
(2004).
\bibitem{Weickert} F. Weickert, R. K\"{u}chler, A. Steppke, L. Pedrero, M. Nicklas, M. Brando, F. 
Steglich, M. Jaime, V.S. Zapf, A. Paduan-Filho, K. A. Al-Hassanieh, C. D. Batista, and P. Sengupta, 
Phys. Rev. B {\bf 85}, 184408 (2012).
\bibitem{papan} N. Papanicolaou, and P. Spathis, J. Phys. C: Solid State Phys. {\bf 20}, L783 
(1987).
\bibitem{Kohama} Y. Kohama, A. V. Sologubenko, N. R. Dilley, V. S. Zapf, M. Jaime, J.A. Mydosh, A. 
Paduan-Filho, K.A. Al-Hassanieh, P. Sengupta, S. Gangadharaiah, A. L. Chernyshev, and C. D. 
Batista, Phys. Rev. Lett. {\bf 106}, 037203 (2011).
\bibitem{mahan} G. D. Mahan, {\it Many-Particle Physics}, 3rd ed. (Kluwer Academic/Plenum 
Publishers, New York, 2000), pp.~177-181.
\bibitem{sakai2005} K. Sakai, and A. Kl\"{u}mper, J. Phys. Soc. Jpn. {\bf 574}, 196 (2005).
\bibitem{louis} K. Louis and C. Gros, Phys. Rev. B {\bf 67}, 224410 (2003).
\bibitem{meisner} F. Heidrich-Meisner, A. Honecker, and W. Brenig, Phys. Rev. B {\bf 71}, 184415 
(2005). 
\bibitem{karadamoglou2004} J. Karadamoglou and X. Zotos, Phys. Rev. Lett. {\bf 93}, 177203 (2004).
\bibitem{zotos1997} X. Zotos, F. Naef, and P. Prelov\v{s}ek, Phys. Rev. B {\bf 55}, 11029 (1997).
\bibitem{long2003} M. W. Long, P. Prelov\v{s}ek, S. El Shawish, J. Karadamoglou, and X. Zotos, 
Phys. Rev. B {\bf 68}, 235106 (2003).
\bibitem{naef1998} F. Naef and X. Zotos, J. Phys.: Condens. Matter {\bf 10}, L183 (1998). 
\bibitem{zotos1999} X. Zotos, Phys. Rev. Lett. {\bf 82}, 1764 (1999).
\bibitem{Sun} X. F. Sun, W. Tao, X. M. Wang, and C. Fan, Phys. Rev. Lett. {\bf 102}, 167202 (2009).
\bibitem{Mukhopadhyay} S. Mukhopadhyay, M. Klanj\v{s}ek, M. S. Grbi\'{c}, R. Blinder, H. Mayaffre, C. Berthier, M. Horvati\'{c}, M. A. Continentino, A. Paduan-Filho, B. Chiari, and O. Piovesana, Phys. Rev. Lett. {\bf 109},  177206 (2012).
\bibitem{Chiatti} O. Chiatti, A. Sytcheva, J. Wosnitza, S. Zherlitsyn, A. A. Zvyagin, V. S. Zapf,
M. Jaime, and A. Paduan-Filho, Phys. Rev. B {\bf 78}, 094406 (2008).
\bibitem{Zvyagin3} S. A. Zvyagin, J. Wosnitza, A. K. Kolezhuk, V. S. Zapf, M. Jaime, A. Paduan-Filho, 
V. N. Glazkov, S. S. Sosin, and A. I. Smirnov, Phys. Rev. B {\bf 77}, 092413 (2008).
\bibitem{kitanine} N. Kitanine, J.M. Maillet, V. Terras, Nucl. Phys. B {\bf 554}, 647 (1999).
\bibitem{caux} J--S. Caux, R. Hagemans and J. M. Maillet, J. Stat. Mech., P09003 (2005).
\bibitem{ovch} A. A. Ovchinnikov, Phys. Lett. {\bf A}377, 3067 (2013).
\bibitem{baxter} R. J. Baxter, J. Stat. Phys. {\bf 108}, no.1/2 (2002).
\bibitem{cox} S. Cox, R. D. McDonald, M. Armanious, P. Sengupta, and A. Paduan-Filho, 
Phys. Rev. Lett. {\bf 101}, 087602 (2008).
\bibitem{AffleckESR} M. Oshikawa and I. Affleck, Phys. Rev. Lett. {\bf 82}, 5136 (1999); Phys. Rev. B {\bf 65}, 134410 (2002);
Y. Maeda, K. Sakai, and M. Oshikawa, Phys. Rev. Lett. {\bf 95}, 037602 (2005); 
M. Brockmann, F. G\"{o}hmann, M. Karbach, A. Kl\'{u}mper, and A. Weisse, \textit{ibid}. {\bf 107}, 017202 (2011). 
\bibitem{maeda} Y. Maeda, and M. Oshikawa, Phys. Rev. B {\bf 67}, 224424, (2003).
\end{thebibliography}
\end{document}